\documentclass[a4paper,iop,numberedappendix]{emulateapj}
\usepackage[colorlinks=true,linkcolor=blue,anchorcolor=blue,citecolor=blue,
urlcolor=blue]{hyperref} 
\usepackage{graphicx}
\usepackage{amssymb,amsmath}
\hypersetup{pdfauthor={Elinor Medezinski}, pdftitle={CLASH: Lensing
    Analysis of the Merging Cluster MACSJ0717}}
\usepackage{natbib}
\newcommand{\simgt}{\lower.5ex\hbox{$\; \buildrel > \over \sim \;$}}
\newcommand{\simlt}{\lower.5ex\hbox{$\; \buildrel < \over \sim \;$}}

\def\btheta{\mbox{\boldmath $\theta$}}

\def\bs{\mbox{\boldmath $s$}}
\def\bp{\mbox{\boldmath $p$}}
\def\bc{\mbox{\boldmath $c$}}
\def\mvir{\mbox{$M_{\rm vir}$}}
\def\mhunit{\mbox{$10^{15}\,M_\sun/h$}}
\def\cvir{\mbox{$c_{\rm vir}$}}
\def\rvir{\mbox{$r_{\rm vir}$}}
\def\mpch{\mbox{${\rm Mpc}/h$}}
\def\kpch{\mbox{${\rm kpc}/h$}}
\def\lcdm{\mbox{$\Lambda$CDM}}
\def\dlsds{\mbox{$D_{\rm ls}/D_{\rm s}$}}
\def\RC{\mbox{$R_{\rm C}$}}

\lefthead{Medezinski et al.}
\righthead{CLASH: MACSJ0717}  

\begin{document}

\title{CLASH: Complete Lensing Analysis of the Largest Cosmic Lens 
MACS\,J0717.5+3745 and Surrounding Structures}

\author{Elinor Medezinski\altaffilmark{1}}      %
\author{Keiichi Umetsu\altaffilmark{2}}         %
\author{Mario Nonino\altaffilmark{3}}           
\author{Julian Merten\altaffilmark{4}}          
\author{Adi Zitrin\altaffilmark{5}}             
\author{Tom Broadhurst\altaffilmark{6,7}}     
\author{Megan Donahue\altaffilmark{8}}         
\author{Jack Sayers\altaffilmark{9}}           
\author{Jean-Claude Waizmann\altaffilmark{10}}     
\author{Anton Koekemoer\altaffilmark{11}}       
\author{Dan Coe\altaffilmark{11}}               
\author{Alberto Molino\altaffilmark{12}}         
\author{Peter Melchior\altaffilmark{13}}        
\author{Tony Mroczkowski\altaffilmark{4,9,14}} 
\author{Nicole Czakon\altaffilmark{9}}         
 \author{Marc Postman\altaffilmark{11}}          
 \author{Massimo Meneghetti\altaffilmark{10}}    
 \author{Doron Lemze\altaffilmark{1}}            
 \author{Holland Ford\altaffilmark{1}}           %
 \author{Claudio Grillo\altaffilmark{15}}         
 \author{Daniel Kelson\altaffilmark{16}}         
 \author{Larry Bradley\altaffilmark{11}}         
\author{John Moustakas\altaffilmark{17}} 
\author{Matthias Bartelmann\altaffilmark{5}} 
\author{Narciso Ben\'itez\altaffilmark{12}}
\author{Andrea Biviano\altaffilmark{3}}         %
\author{Rychard Bouwens\altaffilmark{18}}
\author{Sunil Golwala\altaffilmark{9}}         %
\author{Genevieve Graves\altaffilmark{19}}
\author{Leopoldo Infante\altaffilmark{20}}       %
\author{Yolanda Jim\'enez-Teja\altaffilmark{6}}  
\author{Stephanie Jouvel\altaffilmark{21}}   %
\author{Ofer Lahav\altaffilmark{22}}  
\author{Leonidas Moustakas\altaffilmark{4}}     %
\author{Sara Ogaz\altaffilmark{11}}  
\author{Piero Rosati\altaffilmark{23}}           
\author{Stella Seitz\altaffilmark{24}} 
\author{Wei Zheng\altaffilmark{1}}

\email{elinor@pha.jhu.edu}
\altaffiltext{*}
 {Based in part on data collected at the Subaru Telescope,
  which is operated by the National Astronomical Society of Japan.}
\altaffiltext{1}{Department of Physics and Astronomy, The Johns Hopkins
University, 3400 North Charles Street, Baltimore, MD 21218, USA} 

\altaffiltext{2}{Institute of Astronomy and Astrophysics, Academia
Sinica, P.~O. Box 23-141, Taipei 10617, Taiwan.}  
 \altaffiltext{3}{INAF/Osservatorio Astronomico di Trieste, via
G.B. Tiepolo 11, 34143 Trieste, Italy}
\altaffiltext{4}{Jet Propulsion Laboratory, California Institute of
Technology, MS 169-327, Pasadena, CA 91109, USA} 
\altaffiltext{5}{Universit\"at Heidelberg, Zentrum f\"ur Astronomie, Institut f\"ur Theoretische Astrophysik, Philosophenweg 12, 69120 Heidelberg, Germany} 
\altaffiltext{6}{Dept. of Theoretical Physics and History of Science,
University of the Basque Country UPV/EHU, P.O. Box 644, 48080 Bilbao,
Spain} 
\altaffiltext{7}{Ikerbasque, Basque Foundation for Science, Alameda
Urquijo, 36-5 Plaza Bizkaia 48011, Bilbao, Spain}
\altaffiltext{8}{Department of Physics and Astronomy, Michigan State
University, East Lansing, MI 48824, USA} 
\altaffiltext{9}{Division of Physics, Math, and Astronomy, California
Institute of Technology, Pasadena, CA 91125} 
\altaffiltext{10}{Dipartimento di Astronomia, Universit`a di Bologna, via Ranzani 1, 40127 Bologna, Italy}
\altaffiltext{11}{Space Telescope Science Institute, 3700 San Martin
Drive, Baltimore, MD 21208, USA} 
\altaffiltext{6}{Instituto de Astrof\'isica de Andaluc\'ia (CSIC),
Granada, Spain} 
\altaffiltext{13}{Center for Cosmology and Astro-Particle Physics \&
Department of Physics, The Ohio State University, Columbus, OH, USA} 
\altaffiltext{14}{NASA Einstein Postdoctoral Fellow}
\altaffiltext{15}{Dark Cosmology Centre, Niels Bohr Institute, University of
Copenhagen,
Juliane Mariesvej 30, DK-2100 Copenhagen, Denmark}
\altaffiltext{16}{Observatories of the Carnegie Institution of
Washington, Pasadena, CA, USA} 
\altaffiltext{17}{Department of Physics \& Astronomy, Siena College, 515 Loudon Road, Loudonville, NY, 12211, USA} 
\altaffiltext{18}{Leiden Observatory, Leiden University, Leiden, The 
Netherlands} 
\altaffiltext{19}{Department of Astronomy, University of California, 601 Campbell Hall, Berkeley, CA 94720, USA}
\altaffiltext{20}{Centro de Astro-Ingenier\'ia, Departamento de Astronom\'ia y 
Astrof\'isica, Pontificia Universidad Cat\'olica de Chile,  V. Mackenna 
 4860, Santiago, Chile}
\altaffiltext{21}{Institut de Ci\'encies de l'Espai (IEEC-CSIC), E-08193
Bellaterra (Barcelona), Spain}  
\altaffiltext{22}{Department of Physics and Astronomy, University
College London, London, UK} 
\altaffiltext{23}{ESO-European Southern Observatory, D-85748 Garching bei
M\"unchen, Germany} 
\altaffiltext{24}{Universit\"ats-Sternwarte, M\"unchen, Scheinerstr. 1,
D-81679 M\"unchen. Germany}

\begin{abstract}
  The galaxy cluster MACS\,J0717.5+3745 ($z=0.55$) is the largest known cosmic lens, with complex internal structures seen in deep X-ray, Sunyaev-Zel'dovich effect and dynamical observations. We perform a combined weak and strong lensing analysis with wide-field $BVR_{\rm c}i'z'$ Subaru/Suprime-Cam observations and 16-band {\it Hubble Space Telescope}  observations taken as part of the Cluster Lensing And Supernova survey with Hubble (CLASH).  We find consistent weak distortion and magnification measurements of background galaxies, and combine these signals to construct an optimally estimated radial mass profile of the cluster and its surrounding large-scale structure out to 5\,Mpc$/h$. We find consistency between strong-lensing and weak-lensing in the region where these independent data overlap, $<500\,$kpc$/h$.  The two-dimensional weak-lensing map reveals a clear filamentary structure traced by distinct mass halos. We model the lensing shear field with $9$ halos, including the main cluster, corresponding to mass peaks detected above $2.5\sigma_{\kappa}$. The total mass of the cluster as determined by the different methods is $\mvir\approx(2.8\pm0.4) \times10^{15}\,M_\odot$. Although this is the most massive cluster known at $z>0.5$, in terms of extreme value statistics we  conclude that the mass of MACS\,J0717.5+3745 by itself is not in serious tension with \lcdm, representing only a $\sim2\sigma$ departure above the maximum simulated halo mass at this redshift.

\end{abstract}
 

\section{Introduction} 
\label{sec:intro}

In hierarchical structure formation theories massive clusters are
formed relatively recently and are still growing through the accretion
of substructure. Accretion is predicted to occur preferentially along
filaments with clusters at the nodes of intersection
\citep{1996Natur.380..603B}, a pattern which is now clearly visible in
densely sampled large redshift surveys \citep{Colless+2011_2dF,2005ASPC..329..135H}, all-sky optical surveys such as the Sloan Digital Sky Survey \citep[SDSS;][]{Tegmark04}
and the Baryon Oscillation Spectroscopic Survey \citep[BOSS;][]{2011ApJ...728..126W} and large lensing shear surveys \citep[e.g.,][]{Massey07,VanWaerbeke2013}.
Increasing numbers of
clusters caught in the act of merging are found amongst the most
luminous X-ray sources \citep{Ebeling+2007} and strongest Sunyaev-Zel'dovich effect (SZE)
signal \citep{2011A&A...536A...8P,ACT,Vanderlinde10}. Selection effects are now understood to
strongly favor the detection of gas compressed or shocked during
cluster collision, as illustrated by hydrodynamical simulations \citep{2001ApJ...561..621R,2008ApJ...675.1125B,2012ApJ...748...45M}.

Large simulations of the growth of structure in the context of the
standard cold dark matter ($\Lambda$CDM) \citep{Komatsu+2009_WMAP5} cosmological model have
generated increasingly accurate predictions for the evolution of the
cluster mass function, extending to a limiting halo mass of
approximately $2\times 10^{15}M_\odot$ \citep{Neto07,Duffy08,Zhao+2009,Bhattacharya11}. 
The number density of very massive clusters is predicted to
change relatively rapidly at low redshift, $z<1.0$, where the evolution
is principally sensitive to the cosmological matter density,
$\Omega_m$. To second order one may hope to examine the constancy with
redshift expected for the ``dark-energy'' density
\citep{Allen+2004,Mantz08,Mantz10,SchmidtAllen07} and test for
self-consistency of General Relativity \citep{Rapetti2010}.
Presently, however, there are only indirect determinations of
the masses of statistical samples of clusters, selected by X-ray
means  \citep{2000MNRAS.318..333E,Ebeling+2001_MACS,Vikhlinin+2009_CCC2}, and numbering only less than $\sim300$ clusters, with masses
mostly derived from uncertain X-ray scaling relations \citep{Mantz10b,2006ApJ...640..691V}.  
Although these studies have not yet challenged the
standard model \citep{Mantz10,Vikhlinin+2009_CCC3}, much larger lensing-based surveys of clusters will
eventually provide much more detail, in particular the wide area
Subaru/Hyper Suprime-Cam survey \citep{2010AIPC.1279..120T},
and later Big-BOSS \citep{2009arXiv0904.0468S}, LSST \citep{LSST}, Euclid
\citep{2011arXiv1110.3193L} and WFIRST \citep{2012arXiv1208.4012G}.

Despite the lack of lensing-based cluster mass functions, we may
progress by exploring the most extreme clusters 
\citep{2011ApJ...728...27O,2012MNRAS.420.1754W,2011MNRAS.414.2436C}, in particular the
most distant \citep{2011PhRvD..83j3502H}, because of the exponential sensitivity of the
cluster mass function to the growth of structure
\citep{1995ApJ...447L..81B,Chongchitnan2012}.  Anomalously large masses have been
claimed for clusters at $z\simlt1.5$ \citep{Rosati09,Jee09,Santos11a,Foley11,Tozzi2013}. 
At lower redshifts, accurate full
strong+weak lensing total masses were measured for the massive
clusters such as Abell\,370 ($z=0.375$) and RX\,J1347.5 ($z=0.45$), of
order of $\approx2-3\times10^{15}\,M_\sun$
\citep{Broadhurst08,Umetsu+2011}. These clusters were selected from
all-sky X-ray surveys \citep{2000MNRAS.318..333E,Ebeling+2001_MACS,Ebeling10}, so that although the masses are larger and their
redshifts are lower, the degree of tension with \lcdm\ is not extreme
\citep{2012MNRAS.420.1754W}.
At the present time, no individual cluster has been uncovered that
strains the credibility of the standard \lcdm\ model.

%
Mapping the mass distribution in clusters has provided insight into
the physics that govern dark matter (DM), the main mass component in
the Universe whose nature is largely unknown. The iconic 'Bullet'
cluster \citep{Markevitch02,Clowe04} is an example of a post-merger
cluster collision phase, where the diffuse gas component has been
separated by ram-pressure from the DM and galaxies \citep{Springel07,Mastropietro+Burkert2008}, whereas lensing shows that the DM and the galaxies
are still spatially coincident \citep{Clowe06}.  This system serves to show
that DM is effectively collisionless in nature.  Subsequently, several
new examples of bullet-like clusters have since been discovered with large-scale
supersonic shock fronts \citep{2005ApJ...627..733M,2012ApJ...748....7M,Russell10,2011ApJ...734...10K,2011ApJ...728...82M,2011ApJ...728...27O,Merten11,2012A&A...546A.124V}, which are hard to reconcile
with the expected pairwise velocity distribution of colliding galaxy
clusters, where relative impact velocities in excess of $2000$\,km/s
are unlikely in the context of \lcdm\ \citep{Lee+Komatsu2010,2012MNRAS.419.3560T}.


To shed new light on these mysteries of DM and test
structure formation models with unprecedented precision, the Cluster Lensing And
Supernova survey with Hubble
\citep[CLASH,][]{Postman+2012_CLASH}\footnote{\href{
http://www.stsci.edu/\%7Epostman/CLASH/}{
http://www.stsci.edu/$\sim$postman/CLASH}}, a 524-orbit  {\it Hubble Space
Telescope} ({\it HST}) multi-cycle
treasury program, has been in progress to
couple the lensing power of 25 massive clusters
($M_{\rm vir}=  5-30\times 10^{14}M_\odot$, $\bar{z}_{\rm med}=0.4$) 
with {\it HST} in 16 passbands with full UV/optical/IR
coverage, complemented by Subaru wide-field imaging capabilities
\citep[e.g.,][]{Umetsu+2011_stack,Umetsu+2011}.
Importantly, 20 CLASH clusters were X-ray selected to be relatively
relaxed, in order to examine the concentration-mass relation for a
sample with no strong selection bias toward high concentration
clusters.  A further sample of 5 clusters were selected by their high
lensing magnification properties, with the goal of detecting and
studying high-redshift background galaxies magnified by the cluster
potential.

MACS\,J0717.5+3745 (hereafter, MACSJ0717; $z=0.5458$) is one of the CLASH
high-magnification clusters, and was originally detected by its X-ray
emission, as part of the MAssive Cluster Survey
\citep[MACS,][]{Ebeling+2001_MACS} and independently in the radio
\citep{2003MNRAS.339..913E}. The cluster has the highest X-ray
temperature in the MACS sample \citep[$k_BT_X=11.6$\,keV][ also see
Table~\ref{tab:cluster}]{Ebeling+2007}, and has since been revealed as one of the most dynamically disturbed clusters known. In the core
of this cluster a complex four-component merging activity has been
inferred from optical, X-ray and internal dynamics
\citep{Ma2008,Ma2009}. Low-frequency Radio observation
reveal very complex diffuse radio emission
\citep{2003MNRAS.339..913E,2009A&A...505..991V,Bonafede2009},
indicative of a radio relic or halo, and thought to be a signature of
major mergers and common to many of the distant MACS clusters where
large-scale supersonic shocks are found \citep{2012MNRAS.426...40B}. High
speed gas motion within MACSJ0717 is also tentatively inferred from
dynamical data in a large spectroscopic study, where
structural components were defined on the basis of X-ray emission
peaks \citep{Ma2009}. 
This high relative velocity has been  confirmed in 
the multi-frequency SZE measurements of \cite{2012ApJ...761...47M}, 
which deviate from a thermal 
SZE spectrum in a way that is consistent with kinetic SZE emission.

A strong-lensing (SL) analysis for this cluster has uncovered a complex
elongated tangential critical curve encompassing the central
substructures \citep{Zitrin+2009_0717,Limousin2012}. In terms of the critical area,
this cluster has the
largest strong lensing area, with an equivalent Einstein radius of,
$\theta_E=55\arcsec\pm 3\arcsec$ (at $z_s=2.963$), and an extremely
large mass inside this region, $M_{\rm 2D}(<\theta_E)\sim 7\times10^{14}\,M_\odot$
\citep{Zitrin+2009_0717} was deduced. It has been argued  that
the size of the Einstein radius here may be inconsistent with \lcdm \citep{Zitrin+2009_0717,Meneghetti+2011}.
However, some studies \citep[e.g.,][]{Oguri+Blandford2009,2012A&A...547A..67W} showed that it can be explained as
an extreme case in the context of the lens orientation or other
effects. On the larger scale, MACSJ0717 was shown to be part of a
filamentary structure from galaxy distributions
\citep{2004ApJ...609L..49E}, and from weak-lensing (WL)
\citep{2012MNRAS.426.3369J}, possibly spanning 4\,Mpc in length.

In this paper, we aim to quantify the complex mass properties of
MACSJ0717 and its surrounding large-scale structure (LSS) by employing a
comprehensive weak and strong lensing analysis based on
deep, wide-field Subaru $BVR_{\rm c}i'z'$ imaging, combined with our
recent CLASH {\it HST} imaging. We use our methods to derive a robust
total mass estimate, and calculate meaningful constraints on the
existence of such rare high-mass peaks in a \lcdm\ cosmology.
The paper is organized as follows.  
In Section~\ref{sec:obs}, we describe the observational
dataset, its reduction and WL shape measurements. In
Section~\ref{sec:samples} we describe the selection of cluster and
background galaxies for WL analysis. In Section~\ref{sec:wl-sub} we present the WL analysis using Subaru observation.  In Section~\ref{sec:sl}, we present improved
SL analysis using our new CLASH {\it HST} observations and in Section~\ref{sec:wl-hst} we present a complementary WL analysis using the {\it HST} observations. In Section~\ref{sec:wl1D} we
derive cluster mass profiles from lensing, combining
SL with WL shear and magnification measurements, and  in Section~\ref{sec:wl2D} we present the cluster mass distribution spanning both large scales and zooming in on the core, and constrain individual mass peaks using a multi-halo modeling approach.
In section~\ref{sec:discussion}, we discuss our lensing mass properties, compare with
complementary X-ray and SZE measurements, and contrast the cluster total mass we derive with predictions from \lcdm\ cosmology using extreme value statistics.
Finally, a summary of our work is given in Section~\ref{sec:summary}.

Throughout this paper, we use the AB magnitude system, and
 adopt a concordance \lcdm\ cosmology
with $\Omega_{m}=0.3$, $\Omega_{\Lambda}=0.7$, and
$H_0=100\,h\,$km\,s$^{-1}$\,Mpc$^{-1}$ 
with $h=0.7$.
In this cosmology, $1\arcmin$ corresponds to 268\,kpc\,$h^{-1}=383$\,kpc
 at the cluster redshift, $z=0.5458$.
All quoted errors are 68.3\%
confidence limits (CL) unless otherwise stated. 
The center is taken as the mean location of red-sequence selected cluster members, 
R.A.=07:17:32.63, Dec.=$+$37:44:59.7 (J2000.0).

\section{Subaru+CFHT Observations}
\label{sec:obs}

In this section we present the
data reduction and analysis of MACSJ0717 based on deep Subaru+CFHT
multi-color images (Section~\ref{subsec:data}).  
We briefly describe our WL shape measurement procedure in Section~\ref{sec:shape}.

\subsection{Data Reduction and Photometry}
\label{subsec:data}

\begin{figure*}[!htb]
 \begin{center}
   \includegraphics[width=\textwidth,clip]{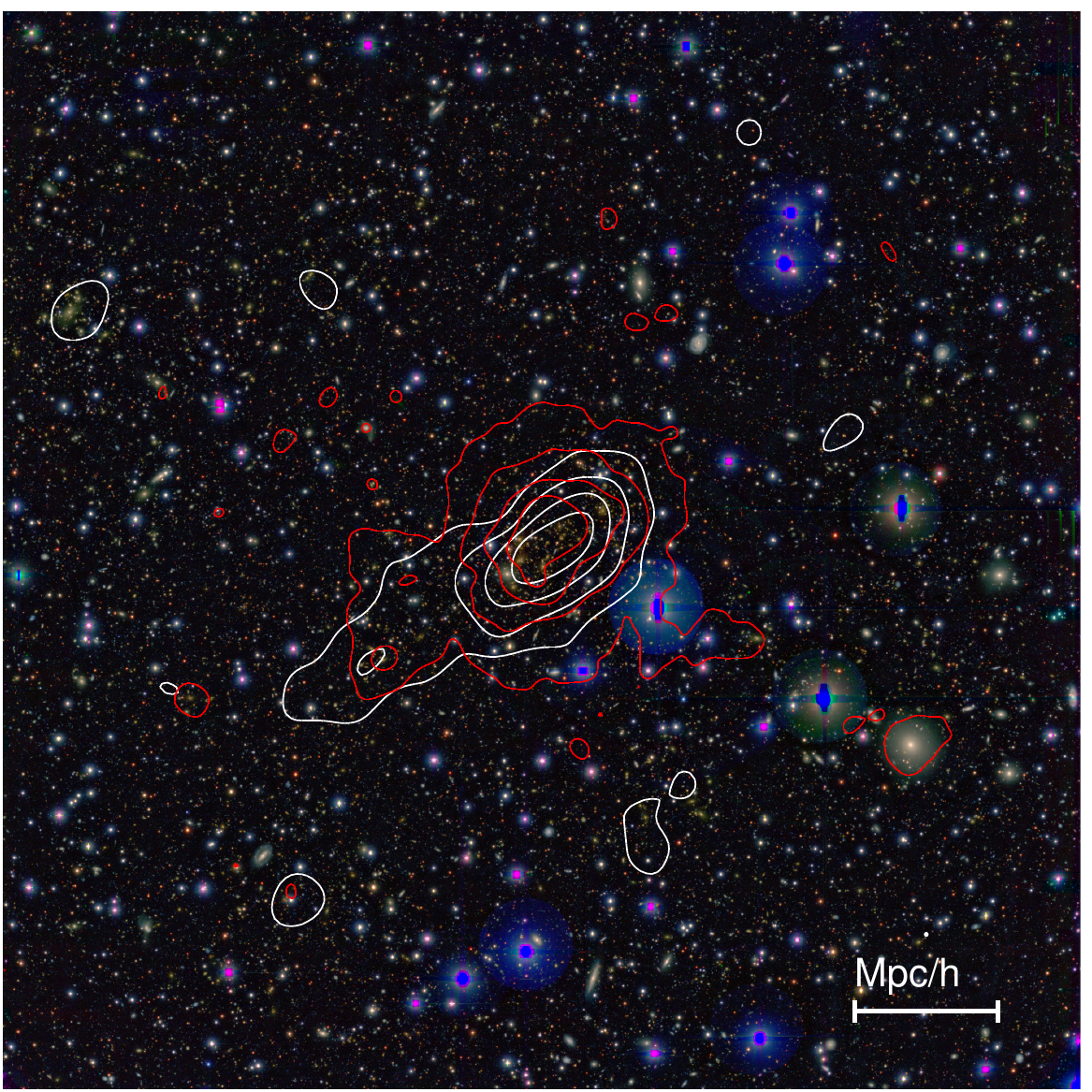}
\caption{
 $28\arcmin\times28\arcmin$ CFHT $u^*$+Subaru $BVR_{\rm c}i'z'$ composite color
image showing the galaxy cluster MACSJ0717 ($z=0.548$). Overlaid are the surface
mass density map reconstructed from our Subaru WL analysis (white contours) and
the X-ray brightness map from XMM-Newton observations (red contours).
North is up and east is to the left. }
\label{fig:color}
 \end{center}
\end{figure*}
We analyze deep $BVR_{\rm c}i'z'$ images of MACSJ0717 
observed with the wide-field camera Suprime-Cam 
\citep[$34^\prime\times 27^\prime$;][]{2002PASJ...54..833M}
at the prime focus of the 8.3-m Subaru
telescope. We observed the cluster on the night of 2010-03-17, in $B,\RC,z'$, to augment shallower observations that existed in
the Subaru archive,
SMOKA\footnote{\href{http://smoka.nao.ac.jp}{http://smoka.nao.ac.jp}}. Some of the archival data for this cluster was taken as part of the ``Weighing the Giants'' program \citep{2012arXiv1208.0597V}.
The seeing FWHMs in the co-added mosaic images are
$0.95\arcsec$ in $B$ (3.84\,ks),
$0.69\arcsec$ in $V$ (2.16\,ks),
$0.79\arcsec$ in $\RC$ (2.22\,ks),
$0.96\arcsec$ in $i'$ (0.45\,ks), and
$0.85\arcsec$ in $z'$ (5.87\,ks)
with $0.20\arcsec$ pixel$^{-1}$, covering a field of approximately
$36\arcmin \times 34\arcmin$.  
The limiting magnitudes are obtained as $B=26.6$, $V=26.4$, $\RC=26.1$,
$i'=25.4$, and $z'=25.6$\,mag for a $3\sigma$ limiting detection within a
$2\arcsec$ diameter aperture.

To improve the accuracy of our photometric redshifts, we also include UV
observations from the Megaprime/MegaCam in the $u^*$-band and Near-IR observations from the WIRCAM
in the $J,K_s$-bands on the CFHT telescope, available from the CFHT archive \footnote{This research used the facilities of the Canadian Astronomy Data Centre
operated by the National Research Council of Canada with the support of the Canadian Space Agency}. Although shallower, and in the case of NIR data the coverage is only of the inner $15\arcmin\times15\arcmin$, these extra bands add important information to help better constrain the SED of galaxies with degenerate fits.

The observation details of MACSJ0717 are listed in
Table~\ref{tab:subaru}.
Figure~\ref{fig:color} shows a $u^*BVR_{\rm c}i'z'$ composite
color image of the cluster central $28\,\arcmin\times28\,\arcmin$, produced automatically using the
publicly available Trilogy software
\citep{Coe2012}.\footnote{\href{http://www.stsci.edu/\%7Edcoe/trilogy/}{
http://www.stsci.edu/$\sim$dcoe/trilogy/}}. We overlay it with the DM map determined from WL (white contours,see Section~\ref{sec:wl2D}) and the smoothed X-ray luminosity map (red contours, see Section~\ref{subsec:xray}). 


Our reduction pipeline derives from \cite{Nonino09} and SDFRED \citep{2004ApJ...611..660O,2002AJ....123...66Y} and has been
optimized for accurate photometry and WL shape measurements. Standard reduction
steps include bias subtraction, flat-field correction (super-flat averaged from
all exposures of the same night where objects have been masked) and
point-spread function (PSF) matching between exposures in the same band (if PSF
variation is evident) as we normally use multi-epoch images taken under
different conditions. Masking of saturated star trails and other artifacts is
then applied.

To obtain an accurate astrometric solution for Subaru observations, we
retrieved processed MegaCam $r$-band (Filter Number: 9601) images from the
CFHT archive and used it as a wide-field reference image.
A source catalog was created from the co-added MegaCam $r$ image, using the
2MASS catalog\footnote{This publication makes use of data products from the Two
Micron All Sky Survey, which is a joint project of the University of
Massachusetts and the Infrared Processing and Analysis Center/California
Institute of Technology, funded by the National Aeronautics and Space
Administration and the National Science Foundation} as an external reference
catalog.  The extracted $r$ catalog has been used as a reference for the SCAMP
software \citep{SCAMP} to derive an astrometric solution for the Suprime-Cam and other CFHT
images. We do not use the CFHT $r$ band image in our photometry as the band is overlapping the Subaru $\RC$ band, but with much lower resolution and depth, and therefore does not carry added information.

For an accurate measure of photometry, we first smear the single exposures of
the same band to the worst seeing. This step is done by using SDFRED/{\sc psfmatch} procedure, after suitable point-like sources have been previously selected. The WL band is chosen to be the band with a combination of best
seeing and deepest observation, $\RC$ in this case (see table~\ref{tab:subaru}). 

Finally, the {\sc Swarp} software \citep{Bertin2002} is utilized to stack the single exposures on a common WCS grid with pixel-scale of $0.2\arcsec$ using the accurate registration that was achieved in the previous step. This assures minimal distortion of the final image. Note, for the
WL band we stack separately data collected at different epochs and different camera rotation angles.

The photometric zero-points for the co-added Suprime-Cam images were
derived from a suitable set of reference stars identified in common with
the calibrated MegaCam data. These zero-points were refined in two independent
ways: first, by comparing with the  {\it HST}/ACS magnitudes of cluster
elliptical-type galaxies, translated to Subaru magnitudes using an elliptical
SED (spectral energy
distribution) template with the BPZ code; subsequently, by fitting SED templates with the BPZ code \citep[Bayesian photometric redshift
estimation,][]{Benitez2000,Benitez+2004} to Subaru photometry of $12$ galaxies
having measured spectroscopic redshifts from the literature\footnote{This research has made use of the NASA/IPAC Extragalactic Database (NED) which is operated by the Jet Propulsion Laboratory, California Institute of Technology, under contract with the National Aeronautics and Space Administration.} \citep{Limousin2012} and calculating model magnitudes. This leads to a final photometric accuracy of $\sim 0.01$\,mag in
all passbands (see also Section~\ref{subsec:depth}).

The eight-band $u^*BVR_{\rm c}i'z'JK_{\rm S}$ photometry catalog was then measured
using SExtractor \citep{1996A&AS..117..393B} in dual-image mode on
PSF-matched images created by ColorPro
\citep{colorpro}, where a combination of $B+V+R+z'$ bands was used as a
deep detection image (we exclude the $i'$ band which is of lesser quality). The stellar PSFs were measured from a combination
of 100 stars per band and modeled using {\sc IRAF} routines.

\subsection{Subaru Shape Measurement}
\label{sec:shape}

For shape measurements, we use our well-tested
WL
analysis pipeline based on the IMCAT package \citep[KSB
hereafter]{1995ApJ...449..460K}, incorporating modifications and
improvements developed and outlined in \citet{Umetsu+2010_CL0024}.
Our KSB+ implementation has been applied extensively to Subaru cluster
observations
\citep[e.g.,][]{Broadhurst05b,Broadhurst08,2007MPLA...22.2099U,UB2008,Okabe+Umetsu2008,
Umetsu+2009,Umetsu+2010_CL0024,Umetsu+2011,Umetsu+2011_stack,Medezinski+2010,
Medezinski+2011,Zitrin+2011_A383,Coe2012,Umetsu+2012,Zitrin2013}. 
Full details of our CLASH WL analysis pipeline are presented in \citet{Umetsu+2012}.

Based on simulated Subaru Suprime-Cam images \citep[see Section 3.2 of][]{Oguri+2012_SGAS,2007MNRAS.376...13M},
we found in our earlier work \citep{Umetsu+2010_CL0024,Umetsu+2012}
that the weak-lensing signal can be recovered
with $|m|\simeq5\%$ of the multiplicative shear calibration bias \citep[as defined by the STEP project: see][]{Heymans06,2007MNRAS.376...13M},  
and $c \sim 10^{-3}$ of the residual shear offset,
which is about one order of magnitude smaller than the typical
distortion signal in cluster outskirts ($|g|\sim10^{-2}$). 
Accordingly, we include in our analysis a calibration factor of $1/0.95$ as
$g_i\to g_i/0.95$
to account for residual calibration.\footnote{Our earlier CLASH weak-lensing work in
\cite{Zitrin+2011_A383}, \cite{Umetsu+2012} and \cite{Coe2012} did not include the 5\% residual correction. Our forthcoming CLASH sample analysis papers will include the 5\% correction factor.}

In this analysis we use the $\RC$-band data taken in 2005 and 2010, which have the best image quality
in our data-sets, taken in fairly good seeing conditions. Two separate co-added
$\RC$-band images are created, one from 2005 (with a total of $450\,$sec,
observer Yasuda) and one from 2010  (with a total of $2160\,$sec, observed by us on
March 2010).
We do not smear the single exposures before stacking (as done for photometric measurements), so as not to degrade and destroy the WL information derived from the shapes of galaxies.
A shape catalog is created for each epoch separately and the catalogs
themselves are then combined by properly weighting
and stacking the calibrated distortion measurements for
galaxies in the overlapping region. The combination of both epochs
increases the number of measured galaxy shapes and improves the statistical
measurement, while not degrading the quality of the shape measurement due to
different seeing and anisotropy at different observing epochs.


\section{Sample Selection}
\label{sec:samples}

For an undiluted WL detection, we need to carefully select a pure sample of
background galaxies. In order to further explore the distribution of
the cluster galaxies, we also identify the cluster members population. We
use the $B,R_{\rm c},z'$ Subaru imaging which spans the full optical wavelength range to perform color-color (CC) selection of cluster and
background samples, as demonstrated by \cite{Medezinski+2010} and detailed below.

\subsection{Cluster Sample Selection}
\label{subsec:clsample}
In Figure~\ref{fig:CC} we show the $B-\RC$ vs $\RC-z'$ distribution of all galaxies
to our limiting magnitude (cyan).
To identify our cluster-dominated area in this CC-space, we also plot up  $B-R_{\rm c}$ vs.
$\RC-z'$ only of galaxies
with small projected distance $R< 3\arcmin$ ($\simlt 1$\,Mpc at
$z_l=0.546$) from the cluster center. A region is then defined according to a
characteristic overdensity in this space (shown as a solid green curve in Figure~\ref{fig:CC}).
Then, {\bf all} galaxies within this distinctive region from the full CC
diagram define the ``green'' sample (green points in Figure~\ref{fig:CC}),
comprising mostly the red-sequence of the cluster and a blue trail of
later-type cluster members. We note that the small overdensity seen bluer (lower-left, $B-\RC, \RC-z'\sim0.7$) than our green sample does not lie at the same redshift of the cluster and is not a bluer population that is part of the cluster, but is in fact comprised of early-type galaxies lying in the foreground of the cluster, at about $z\sim0.33$, which we will discuss further as part of our multi-halo mass modeling (Section~\ref{subsec:2Dhalo}).

The number density profile of the green sample is steeply rising toward the center
(Figure~\ref{fig:nplot}, green crosses). The low number density at large clustercentric radius
is indicative of negligible contamination of background galaxies of this sample.
The WL signal for
this population is found to be consistent with zero at all radii (Figure
\ref{fig:gt1D}, green crosses), also indicating the reliability of our
procedure.
For this population of galaxies, we find a mean photometric redshift of
$\langle z_{\rm phot}\rangle \approx 0.56$ (see Section
\ref{subsec:depth}), consistent with the cluster redshift.
Importantly, the green sample marks the region that contains a majority
of unlensed galaxies, relative to which we select our background
samples, as summarized below.

\subsection{Background Sample Selection}
\label{subsec:color}

\begin{figure}[tb]
 \begin{center}
  \includegraphics[width=0.5\textwidth,clip]{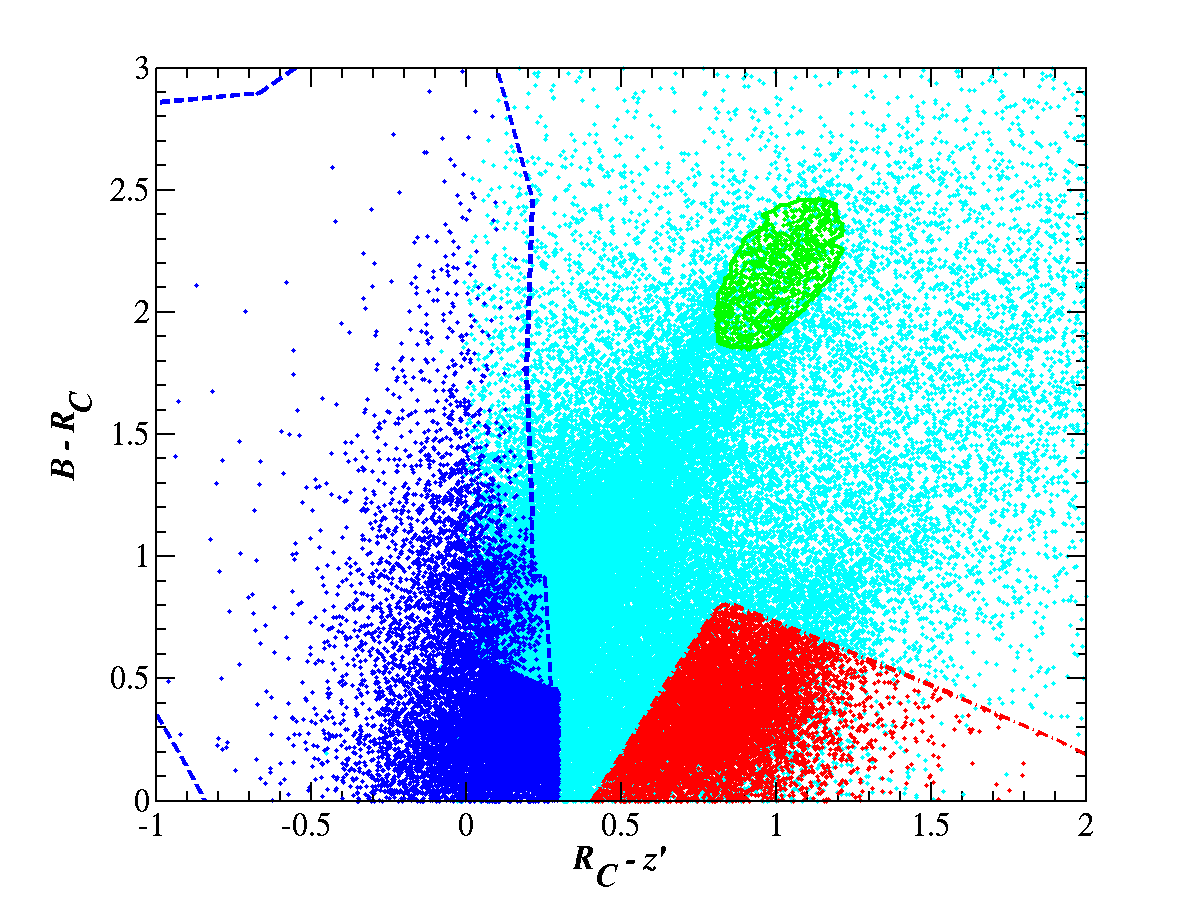}
 \end{center}
\caption{
``Blue'' and ``red'' background galaxies are selected for WL analysis (lower
left blue dashed and right red dot-dashed contours, respectively) based on
Subaru $B,\RC,z'$ color-color-magnitude selection. All galaxies (cyan) are
shown in the diagram. At small clustercentric radius, an overdensity of cluster
galaxies defines our ``green'' sample (green solid contour), comprising mostly the
red sequence of the cluster and some blue trail of later type cluster members.
The background samples are well isolated from the green region and satisfy other
criteria as discussed in Section~\ref{subsec:color}.}
\label{fig:CC}
\end{figure}

A careful background selection is critical for a WL analysis
so that unlensed cluster members and foreground galaxies do not dilute
the true lensing signal of the background
\citep{Broadhurst05b,Medezinski+07,UB2008,Medezinski+2010}.
This dilution effect is simply to reduce the strength of the lensing
signal when averaged over a local ensemble of galaxies \citep[by a
factor of 2--5 at $R\simlt 400\,{\rm kpc}\,h^{-1}$; see 
Figure 1 of][]{Broadhurst05b}, particularly at small radius
where the cluster is relatively dense, in proportion to the fraction of
unlensed galaxies whose orientations are randomly distributed. 

We use the background selection method of \citet{Medezinski+2010} to
define undiluted samples of background galaxies,
which relies on empirical correlations for galaxies in
color-color-magnitude space derived from the deep Subaru photometry, by 
reference to evolutionary tracks of galaxies
\citep[for details, see][]{Medezinski+2010,Umetsu+2010_CL0024} as well
as to the deep photometric-redshift survey in the COSMOS field
\citep{Ilbert+2009_COSMOS}.  

For MACSJ0717, we have a wide wavelength coverage 
($BVR_{\rm c}i'z'$) of Subaru/Suprime-Cam.
We therefore make use of the $(B-R_{\rm c})$ vs. $(R_{\rm c}-z')$
CC-diagram to carefully select two distinct background
populations which encompass the red and blue branches of galaxies.
We limit the red sample to $z'<25$\,mag in the reddest band, corresponding
approximately  to a $5\sigma$ limiting magnitude within $2\arcsec$
diameter aperture.  We extend the magnitude limit of the blue samples further to $z'<26$\,mag, where the number density of galaxies grows significantly higher, especially for bluer galaxies whose faint-end slope of the luminosity function is rising, giving a much improved WL statistical measurement.

For the background samples, we define conservative color limits, 
where no evidence of dilution of the WL signal is visible, to
safely avoid contamination by unlensed cluster members and foreground
galaxies. 
The color boundaries of our ``blue'' and ``red'' background samples are shown
in Figure \ref{fig:CC}.
For both the blue and red samples, we find a consistent, rising
WL signal (see Section~\ref{subsec:gt}) all the way to the center of the cluster, as shown
in Figure \ref{fig:gt1D}.

As a further consistency check, we also plot in Figure~\ref{fig:nplot} the
galaxy surface number density as a function of clustercentric radius, $n(\theta)$, for the blue
and red samples. As can be seen, no clustering is observed toward the center for
the background samples, which demonstrates that there is no significant
contamination by cluster members in these samples.  The red sample
systematically decreases in projected number density toward the cluster center,
caused by the lensing magnification effect.
A more quantitative magnification analysis is given in Section
\ref{subsec:magbias}.

To summarize, our CC-selection criteria yielded a total of 
N=10490, 1252, and 11998 galaxies, for the red, green, and blue
photometry samples, respectively (see Table~\ref{tab:color}).
For our WL distortion analysis, 
we have a subset of 4856 and 4738 galaxies in the red and blue samples
(with usable $\RC$ shape measurements), respectively (see Table~\ref{tab:wlsamples}). 

\subsection{Depth Estimation}
\label{subsec:depth}

\begin{figure}[tb]
 \centering
\includegraphics[width=0.5\textwidth]{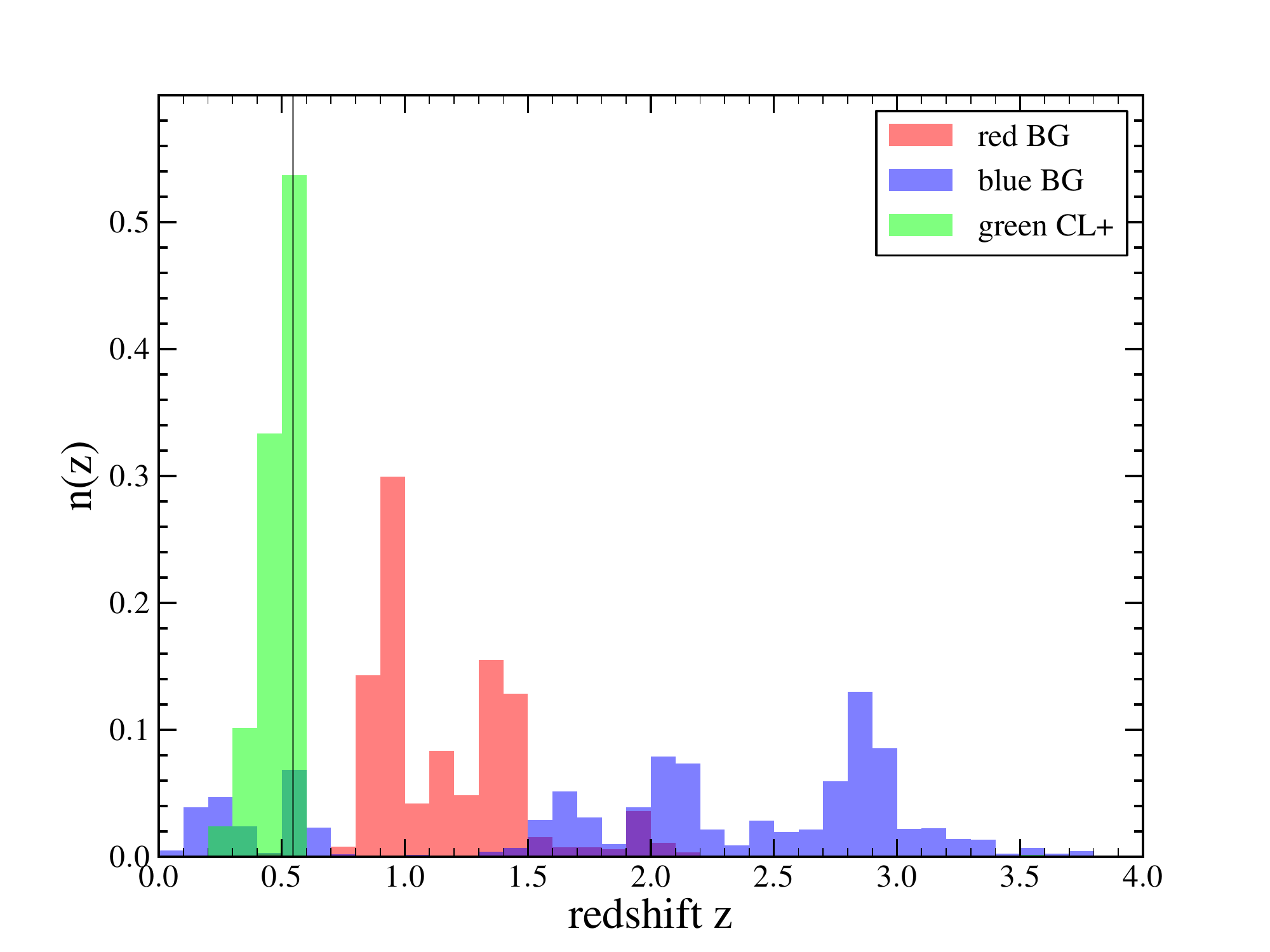}
\caption{Redshift distributions of CC-selected green, red and blue samples using the BPZ photo-z's based on Subaru+CFHT imaging. The cluster redshift is marked with a black line. }
\label{fig:zdist}
\end{figure}


\begin{figure}[tb]
 \begin{center}
\includegraphics[width=0.5\textwidth,clip]{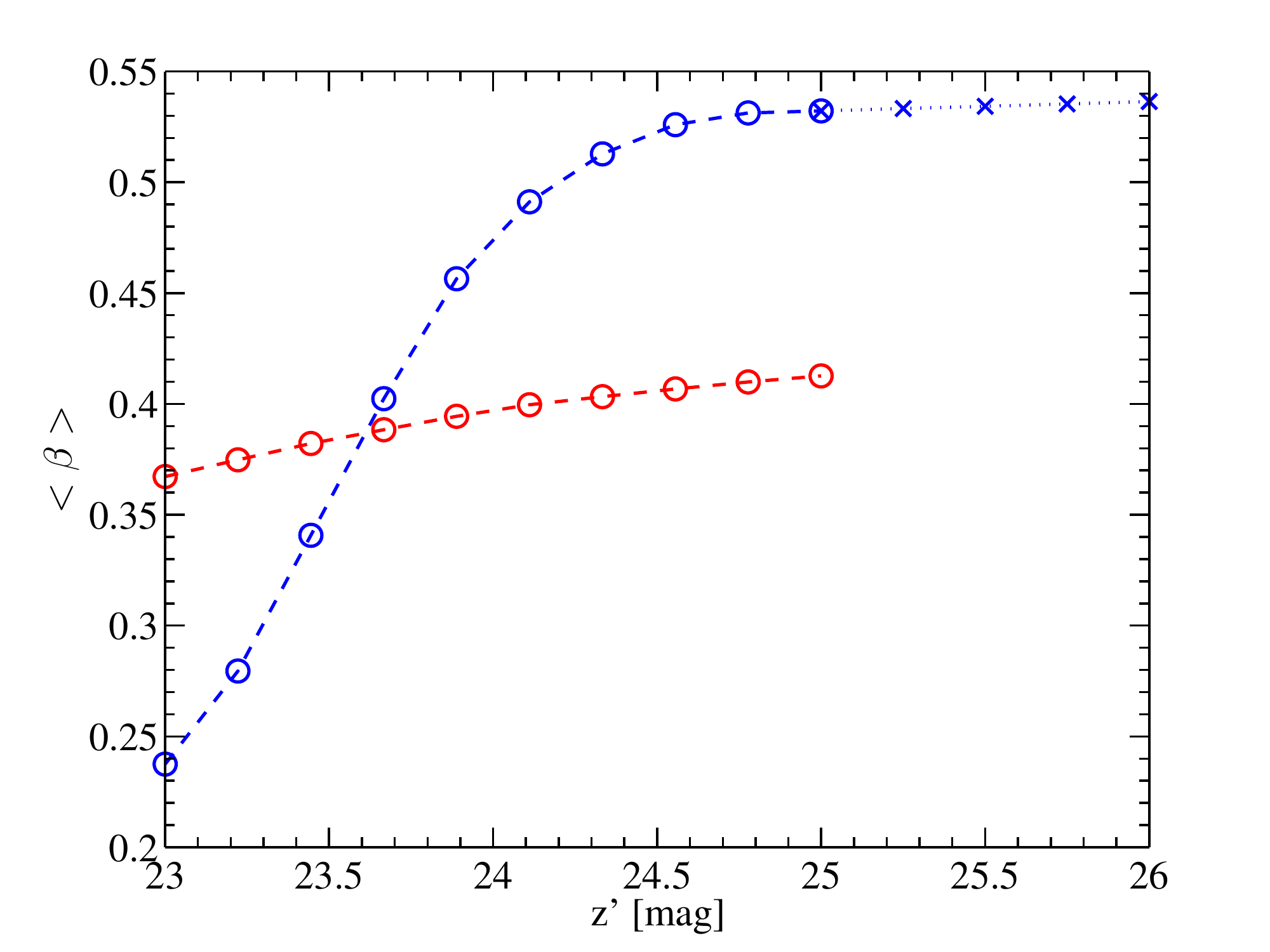}
 \end{center}
\caption{Lensing depth ($\dlsds$) as a function of Subaru $z'$-band limiting magnitude for the red and blue background samples, as estimated from the photometric redshifts of COSMOS to $z'<25$ (circles). In order to estimate the depth of the blue sample to its limiting magnitude of $z'<26$ we extrapolate the curve (x's).}
\label{fig:D-mag}
\end{figure}

The lensing signal depends on the source redshift $z_s$ 
through the distance ratio $\beta(z_s)=\dlsds$, where $D_{\rm ls},D_{\rm s}$ are the angular diameter distances between the lens and the source, and the observer and the source, respectively.
We thus need to estimate and correct for the respective depths 
$\langle\beta \rangle$ of
the different galaxy samples, when converting the
observed lensing signal into physical mass units.

For this we utilize BPZ to measure photometric
redshifts (photo-$z$s) $z_{\rm phot}$ using our  
deep Subaru+CFHT $u^*BVR_{\rm c}i'z'JK_{\rm s}$ photometry (Section
\ref{subsec:data}).
BPZ employs a Bayesian inference  where the redshift likelihood is weighted
by a prior probability, which yields the probability density $P(z,T|m)$ of a
galaxy with apparent magnitude $m$ of having certain redshift $z$ and
spectral type $T$.
In this work we used a new library (N. Benitez 2012, in prep.) composed of 10
SED templates originally from PEGASE \citep{Fioc+Rocca1997} but re-calibrated
using the FIREWORKS photometry and spectroscopic redshifts from
\citet{Wuyts+2008} to optimize its performance.  This  library includes five
templates for elliptical galaxies, two for spiral galaxies, and three for
starburst galaxies.  
In our depth estimation we utilize BPZ's ODDS parameter, which measures the
amount of probability enclosed within a certain interval $\Delta z$ centered on
the primary peak of the redshift probability density function (PDF), serving as
a useful measure to quantify the reliability of photo-$z$ estimates.

%
We only consider galaxies from the WL-matched catalogs, as those are the galaxies
from which we estimate the lensing signal and finally the mass profile. To make this estimate more robust, we use galaxies for which the photo-z was determined using all available 8 bands.
We show the normalized redshift distribution of these galaxies in each of the
green, red and blue samples in Figure~\ref{fig:zdist}.
Still, since only 12 spectroscopic redshifts are publicly available in this
field, it is difficult to estimate the reliability of our photo-z's. We therefore
further compare our results with the more reliably estimated depths we derive from the COSMOS catalog
\citep{Ilbert+2009_COSMOS}, which has robust photometry and photo-$z$
measurements for the majority of galaxies with $i'<25$\,mag. For each sample, we
apply the same CC selection to the COSMOS photometry 
and obtain the redshift distribution $N(z)$ of field galaxies. Since COSMOS is
only complete to $i'<25$\,mag, we derive the mean depth as a function of
magnitude (fig.~\ref{fig:D-mag}) up to that limit, and extrapolate to our sample
limiting magnitude, $z'=25$ in the case of the red sample, and $z'=26$ in the
case of the blue sample.


For each background population,
we calculate a weighted mean of the distance ratio $\beta$ (mean lensing depth) as
\begin{equation}
\label{eq:beta}
\langle \beta \rangle =
  \frac{\int\!dz\,w(z) N(z)\beta(z)}{\int\!dz\,w(z)N(z)},
\end{equation}
where $w(z)$ is a weight factor; $w$ is taken to be the Bayesian ODDS
parameter for the BPZ method, and $w=1$ otherwise.
The sample mean redshift $\langle z_s\rangle$ is defined similarly to 
Equation (\ref{eq:beta}).

In Table \ref{tab:wlsamples} we summarize the mean depths 
$\langle\beta \rangle$ and the effective source redshifts 
$z_{s,{\rm eff}}$ for our background samples.
For each background sample,
we obtained consistent mean-depth estimates 
$\langle \beta\rangle$  (within $2\%$)
using the BPZ- and COSMOS-based methods.
In the present work, we adopt a conservative uncertainty of $5\%$ in the
mean depth for the combined blue and red sample of
background galaxies, 
$\langle\beta({\rm back})\rangle=0.48\pm 0.03$, which corresponds 
to $z_{s,{\rm eff}}=1.26\pm 0.1$.   We marginalize over this
uncertainty when fitting parametrized mass models to our WL data.

\section{Subaru Weak-Lensing Analysis}
\label{sec:wl-sub}


In this section we describe the WL analysis based on our deep Subaru imaging data.
In  Section~\ref{subsec:gt} we derive cluster lens distortion and in Section~\ref{subsec:magbias} we derive the
magnification radial profiles from the data. 

\subsection{Tangential Distortion Analysis}
\label{subsec:gt}

The shape distortion of an object is described by the complex
reduced-shear, $g=g_1+ig_2$,
where the reduced-shear is defined as  \citep[in the subcritical regime; see, e.g.,][]{2001PhR...340..291B},
\begin{equation}
\label{eq:g}
g_{\alpha}\equiv\gamma_{\alpha}/(1-\kappa),
\end{equation}
where $\gamma$ is the complex gravitational shear field and is non-locally related to the convergence, $\kappa = \Sigma/\Sigma_{\rm crit}$, which is the surface mass
density in units of the critical surface-mass density for lensing, $\Sigma_{\rm crit} = \frac{c^2}{4\pi GD_{\rm l}}\beta^{-1}$.
The tangential component of the
reduced-shear, $g_{+}$, is used to obtain the azimuthally-averaged
distortion due to lensing, and computed from the distortion
coefficients $(g_{1},g_{2})$:
\begin{equation}
g_{+}=-( g_{1}\cos2\theta +  g_{2}\sin2\theta),
\end{equation}
where $\theta$ is the position angle of an object with respect to the
cluster center. The uncertainty in the object $g_{+}$ measurement is
$\sigma_{+} = \sigma_{g}/\sqrt{2}\equiv \sigma$ in terms of the RMS
error $\sigma_{g}$ for the complex reduced-shear measurement.  For each galaxy, $\sigma_{g}$ is the variance for the reduced shear estimate computed from $50$ neighbors identified in the $r_g$--$\RC$ plane.
To improve the statistical significance of the distortion measurement, we
calculate the weighted average of $g_{+}$ 
\begin{equation} \label{eq:gt}
g_{+,i} \equiv g_+(\theta_i)=
\left[
\displaystyle\sum_{k\in i} w_{(k)}\, g_{+(k)}
\right]
\left[
\displaystyle\sum_{k\in i} w_{(k)}\right]^{-1},
\end{equation}
where the index $k$ runs over all objects located within the $i$th
radial bin with a weighted center of $\theta_i$,
and  $w_{(k)}$ is the weight for the $k$-th object,
\begin{equation}
\label{eq:gtw}
w_{(k)}=1/(\sigma_{g(k)}^2+\alpha^2), 
\end{equation}
where $\alpha^2$ is the
softening constant variance. We choose $\alpha=0.4$, which is a
typical value of the mean RMS $\bar{\sigma}_g$ over the background
sample.
The uncertainty in $g_{+,i}$ is calculated 
from a bootstrap error analysis \citep[for details, see][]{Umetsu+2012}. 
Since WL only induces
curl-free tangential distortions, the $45\deg$-rotated component, $g_{\times} = -(g_2 \cos 2\phi - g_1\sin 2\phi)$, is expected to vanish. It is therefore useful as a check for systematic errors.

\begin{figure}[tb]
\includegraphics[width=0.5\textwidth]{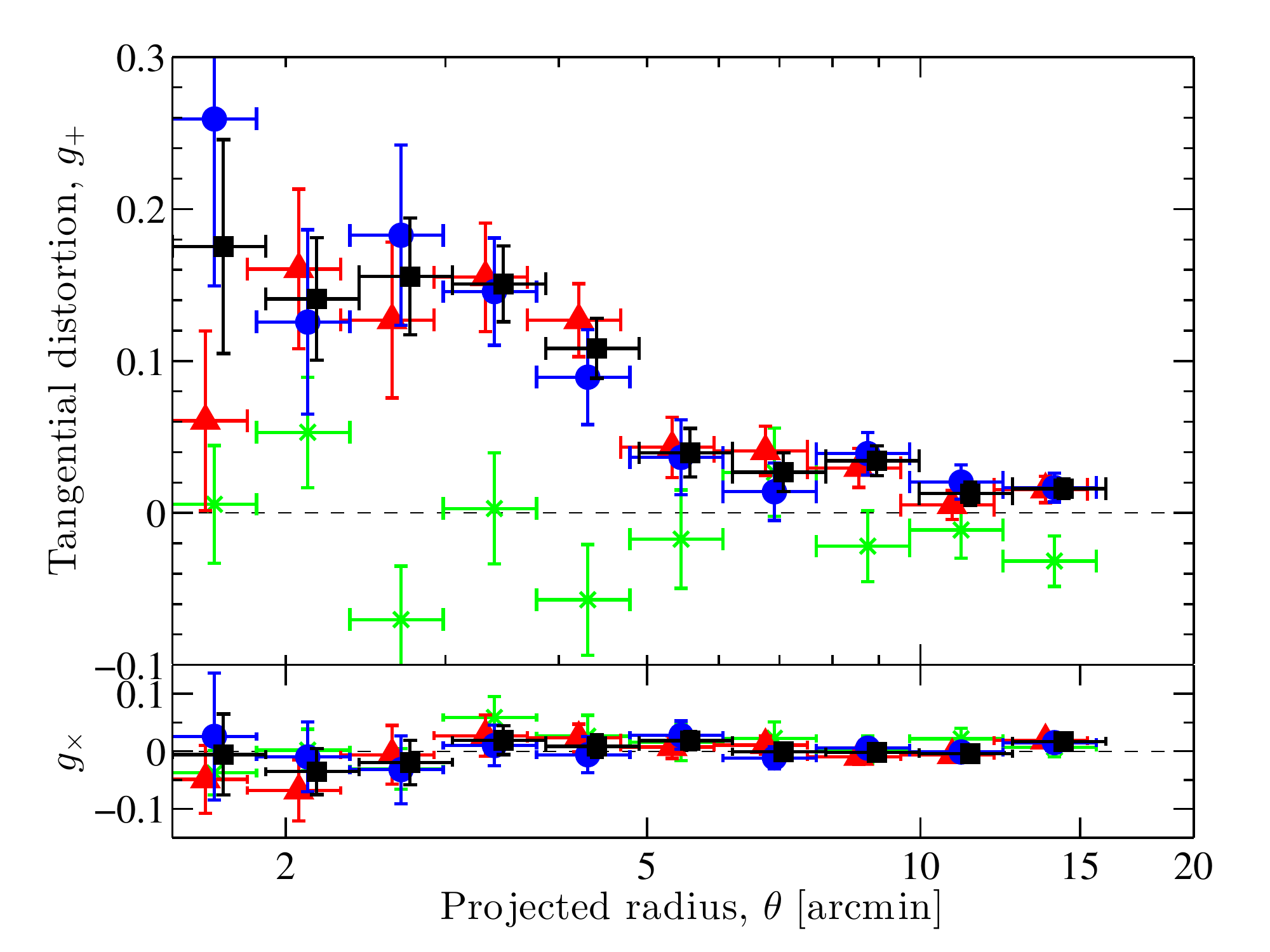}
\caption{
Azimuthally-averaged radial profiles of the tangential reduced-shear $g_+$
(upper panel) and the $45^\circ$ rotated ($\times$) component $g_{\times}$
(lower panel) for our  red (triangles), blue (circles), green (crosses), and
blue+red (squares) galaxy samples. The symbols for the red and blue samples are horizontally shifted
for visual clarity. For all of the samples, the $\times$-component is consistent
with a null signal detection well within $2\sigma$ at all radii, indicating the
reliability of our distortion analysis.}
\label{fig:gt1D}
\end{figure}

In Fig.~\ref{fig:gt1D} we plot the radial profile of $g_+$ of the green,
red and blue samples defined above. The black points represent the
red+blue combined sample (also shown in Figure~\ref{fig:wldata}, upper panel), showing the best estimate of the lensing
signal, which is detected at $11.6\sigma$ significance over the full radial range.
The red and blue samples profiles rise continuously toward
the center of the cluster, and agree with each other within the errors, except at the very central bin, $\theta\simlt2\arcmin$, where measurement are approching the non-linear regime, given the extremely elliptical shape of the tangential critical curve \citep[see Figure 1 of][]{Zitrin+2009_0717}.
The overall rising trend and agreement demonstrate that both the red and blue samples are dominated by background
galaxies and are not contaminated by the cluster at all radii.  The
$g_+$ profile of the green sample agrees with zero at all radii.
The measured zero level of tangential distortion reinforces our CC selection of the green sample to consist of mostly cluster members.

\subsection{Magnification-Bias Analysis}
\label{subsec:magbias}

\begin{figure}[tb]
 \centering
\includegraphics[width=0.5\textwidth]{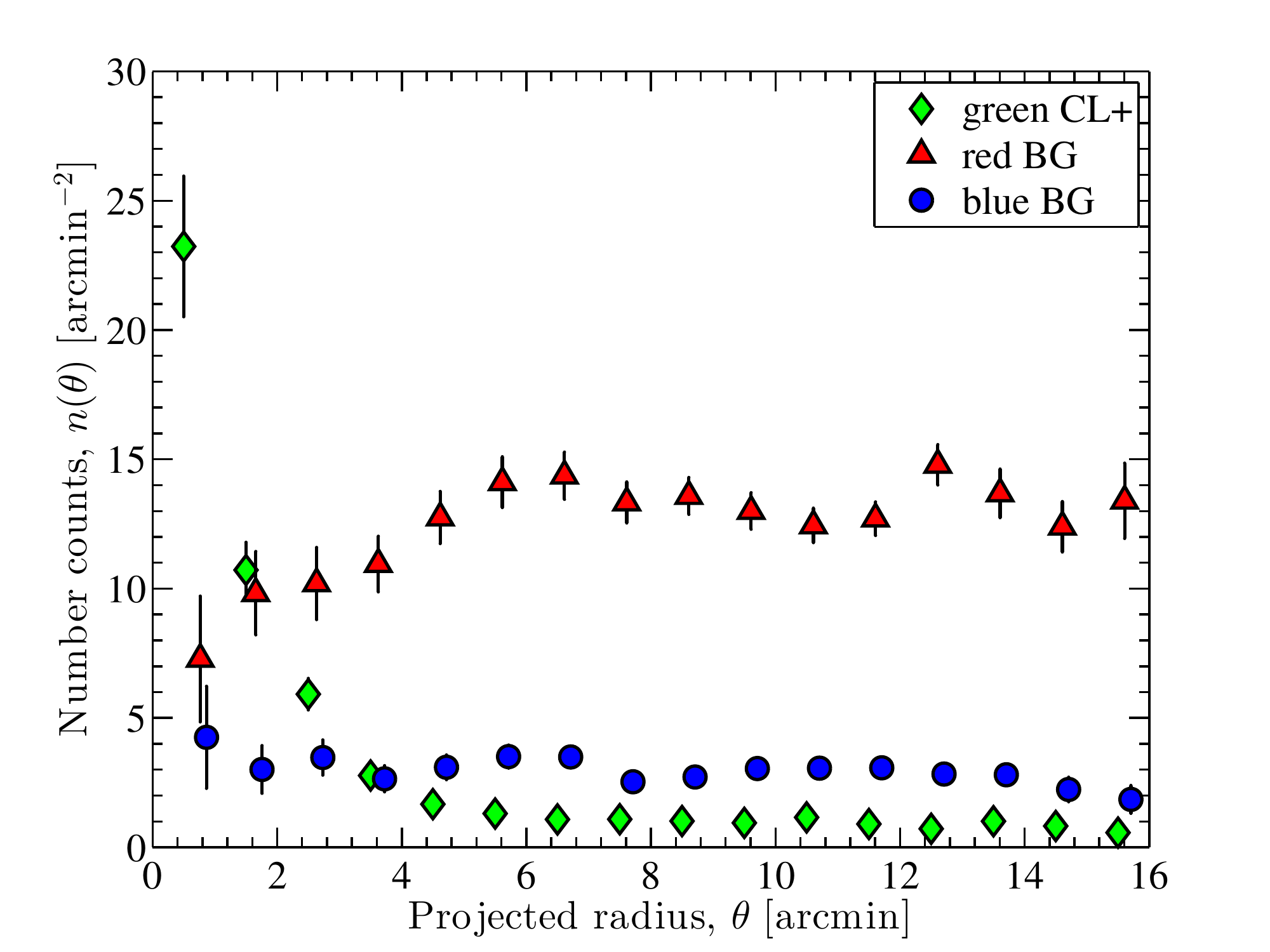}
\caption{Surface number density profiles $n(\theta)$ of Subaru
$BR_{\rm c}z'$-selected samples of galaxies.  The red (triangles) and blue
(circles) samples comprise background galaxies, and the green (crosses) sample comprises mostly cluster galaxies, as is evident by their steeply rising number counts toward the center.
See also Figure \ref{fig:wldata}.}
\label{fig:nplot}

\end{figure}

We follow the prescription of \citet{Umetsu+2011} to measure the magnification bias signal as a function of distance from the cluster center, which depends on the
intrinsic slope of the luminosity function of background sources $s$, as
\begin{equation} 
\label{eq:magbias}
n_\mu(\btheta)=n_0\mu(\btheta)^{2.5s-1},
\end{equation}   
where $n_0 = dN_0(<m_{\rm cut})/d\Omega$ is the unlensed mean
number density of background sources for a 
given magnitude cutoff $m_{\rm cut}$, approximated locally as a
power-law cut with slope,
$s=d\log_{10} N_0(<m)/dm >0$.
We use the red sample of background galaxies (Section \ref{subsec:color}), for
which the intrinsic count slope $s$
at faint magnitudes is relatively flat, $s\sim 0.1$, so that a net
count depletion results \citep{Broadhurst05b,UB2008,Umetsu+2010_CL0024,Umetsu+2011}.
In contrast, the blue background population has
a steeper intrinsic count slope close to the lensing invariant slope
($s=0.4$).



Here we use the approach developed in \citet{Umetsu+2011} to
measure the azimuthally-averaged surface number density profile 
of the red galaxy counts, $n_{\mu,i}\equiv n_\mu(\theta_i)$ (red triangles in Fig.~\ref{fig:nplot}),
taking into account and correcting for masking of background galaxies 
due to bright cluster galaxies, foreground objects, and
saturated objects.
The errors $\sigma_{\mu,i}$ for $n_{\mu,i}$
include both contributions from Poisson errors in the counts
and contamination by intrinsic clustering of red background
galaxies.
The normalization and slope parameters for the red sample are reliably
estimated 
outside the lensed region, 
by virtue of the wide-field imaging with Subaru/Suprime-Cam. We find  
$n_0=13.3\pm 0.3$\,galaxies/acrmin$^{2}$
and
$s=0.123\pm 0.048$.

We show in the bottom panel of Figure \ref{fig:wldata} the 
measured magnification profile from our flux-limited sample of red
background galaxies ($z'<25$\,mag; see Table \ref{tab:color}) with and without the masking correction applied (red circles and green crosses, respectively).
A clear depletion of the red counts is seen in the central,
high-density region of the cluster and detected out to $\simlt
4\arcmin$ from the cluster center.
The radially integrated significance of the detection of the depletion
signal is $5.3\sigma$.

The high-level WL shear and magnification profile we derived here can be combined together to reconstruct the underlying mass profile. However, to better resolve the core of the profile, we require further constraints which are enabled by SL. We therefore derive the SL mass profile in the next section.

\section{{\it HST} Strong Lensing Analysis}
\label{sec:sl}

For a massive cluster, the weak- and strong-regimes contribute
similar logarithmic coverage of the mass profile.
Hence, the central SL information is crucial
in a cluster lensing analysis \citep{Umetsu+2011,Umetsu+2011_stack,Umetsu+2012}. In this section we derive a SL model to compare with our WL profile in the overlap region, and to combine with WL in deriving the mass reconstruction in section~\ref{sec:wl1D}.

First we summarize our well-tested approach to strong-lens
modeling, developed by \citet{2005ApJ...621...53B} and optimized further by
\citet{Zitrin+2009_CL0024} \citep[see also][]{Zitrin2013}.
Briefly, the adopted parametrization is as follows. Cluster members, chosen by a F814W-F555W color criterion, are each represented by a power-law mass density profile. The superposition of all galaxy contributions constitutes the galaxy, lumpy component for the model. This component is then smoothed using a 2D spline interpolation to comprise the DM component. The two components are then added with a relative weight. In order to allow further degrees of freedom (DOF), and higher effective ellipticity of the critical curves, an external shear is added. In total, the method thus includes six basic free parameters: (1) the power-law of the galaxy mass profile; (2) the smoothing (polynomial) degree of the DM component; (3) the relative weight of the galaxy to the DM component; (4) the overall scaling or normalization; (5) the amplitude and (6) angle of the external shear \citep[for more details see][]{Zitrin+2009_CL0024}. 

Using this method, in \cite{Zitrin+2009_0717} we performed the first SL analysis of this cluster using 3-band publicly available {\it HST} imaging, and found 34 multiple-images from 13 lensed sources, uncovering that MACSJ0717 is the largest known lens. 
As part of CLASH \citep{Postman+2012_CLASH},  we further observed MACSJ0717 with {\it HST} in 16 filters with the  ACS and WFC3/UVIS+IR cameras.
We thus revise here our primary multiple image identification, and follow in general the multiple image sets listed in Table 1 of \citet[][hereafter L12]{Limousin2012}, who recently used \cite{Zitrin+2009_0717}'s systems (with the exclusion of three systems, 9-11) and added {\bf five} additional multiple systems which were  sunsequently verified with the help of the publicly-available CLASH WFC3/IR images.
%
Although we agree with most of the identifications and revisions by L12, we determine image 1.5 to be at a different location, RA=07:17:37.393, Dec=+37:45:40.90. We also confirm this new location significantly improves the SL parametric solution of L12 (M. Limousin, private communication).
In addition, although the model suggests these may be counter images of the same source, we omit system 2 from our list, since it strongly deviates from the $\dlsds$ scaling relation expected, and as determined from visual inspection. Aside from these corrections, we use in total 43 multiple-images coming from the other 14 sources listed in L12 as our SL constraints here.
Internal bright knots in some of the images are added as further constraints.  We note that since we have sufficient information to constrain the mass model, we do not attempt to find additional multiple images in the current paper, and we leave this for future work,  mainly in the framework of the upcoming Frontier Fields program.

We use a several dozen thousand step Monte-Carlo Markov-Chain (MCMC) minimization in order to find the best-fit solution, defined by the image-plane reproduction $\chi^{2}$.
The advantage of this light-traces-mass \citep{Zitrin+2009_0717} method is that even very complex systems such as MACSJ0717 are still well fitted by this simple procedure, although it may not be expected to yield an RMS as low as in a multi-halo/parameteric fit (e.g. L12).
Note however, that even with a somewhat higher RMS, the representation is still highly credible as it allowed the identification of many multiple image systems \citep{Zitrin+2009_0717}.

Here, in practice, to allow for more freedom and a better RMS, we also leave the relative weight of 10 galaxies to be freely optimized by the MCMC. With this, the final model we present here has an RMS of $3.86\arcsec$ and a $\chi^{2}$ of $\simeq 334$ (with a location error of $\sigma=1.4\arcsec$), over 32 DOF. The relatively large reduced $\chi^2$ (compared with the reasonable RMS) may indicate that in such a complex system, a position error of $1.4\arcsec$ may be underestimated, not taking into account stronger LSS and complexity effects. Since the best-fit statistical solution defined by the image positions may not always reproduce the multiple images with the right internal shape or orientation, multiple images are then sent to the source plane and back through the lens to test, by eye, the reproduction of other multiple images of the same systems \citep[e.g.,][]{Zitrin+2009_0717,Zitrin+2009_CL0024}. Note that only 5 systems have spectroscopic redshift, so that we use the input from L12 as the predicted redshift of the other systems. We do not leave any of these redshifts to be optimized by the model, which may have lowered further the RMS of our model.

We present the azimuthally-averaged projected mass density profile from the resulting SL model of MACSJ0717 in Figure~\ref{fig:1d-2d} (green curve). The density profile shows a remarkably flat core out to $\simlt180\kpch$, as was noted in previous SL analyses of this cluster \citep{Zitrin+2009_0717}.
The shallow profile is in accordance with the non-relaxed appearance of this cluster and reported multiple mass clumps at the core of the cluster. 
According to our SL model, the total projected mass enclosed within a radius of $60\pm6\arcsec$, the effective Einstein radius at $z_s=2.963$, is $M_{\rm 2D}(<60\arcsec)=(4.87 \pm 0.35) \times 10^{14}\,M_\sun/h$, which is in good agreement with the mass enclosed within the tangential critical curve, $M_{\rm 2D}=(5.5\pm 0.35)\times 10^{14}\,M_\sun/h$. 
We now turn to the WL analysis in the next sections in order to recover the mass over the full scale of the cluster. 

\section{{\it HST} Weak-Lensing Analysis}
\label{sec:wl-hst}

In order to further constrain the mass profile in the cluster center, where
the small Subaru number density leads to large uncertainties, we preform a complementary WL analysis of the {\it HST} 16-band data in the weak regime.
Details of our reduction, photometry and
photo-z pipeline were given in previous papers \citep{Postman+2012_CLASH,Zitrin+2012_M1206,Coe2012}.
Here we further produced specialized drizzled images, optimized for WL, consisting of drizzling each visit in the ``unrotated'' frame of the ACS/WFC3 detectors, using a modified version of the "Mosaicdrizzle" pipeline \cite[described more fully in][]{2011ApJS..197...36K}. This allows accurate PSF treatment that does not compromise the intrinsic shape measurements required by WL pipelines.
The RRG \citep{2000ApJ...536...79R} WL shape measurement package  was then used to measure shapes in each
of six ACS bands (F435W, F475W, F625W, F775W, F814W and F850LP), and the DEIMOS \citep{2011MNRAS.412.1552M} package  was used to measure shapes in the WFC3/IR F160W band. We exclude
objects with $S/N<10$ and ${\rm size}<0.1\arcmin$. All the shape catalogs were then
matched to the deep multi-band photometric catalog and merged where there was more than one
shape measurement per object, using its S/N as input weight of each measurement, according
to eq.~\ref{eq:gtw} with $\alpha=0.2$ as the softening kernel in this case.

Since the {\it HST} field has reliable photometric redshifts measured from 16 bands spanning UV to
IR, here we rely on those for a secure selection of background galaxies. We
define our background sample as galaxies having $0.8<z_{\rm b}<4,\, z_{\rm b,min}>0.6,\, z_{\rm b,max}<5,\,22.5<m_{F625W}<27.5$, where $z_{\rm b}$ is the bayesian photometric redshift from BPZ, and $z_{\rm b,min},z_{\rm b,max}$ are the 68\% lower and upper bound on the photometric redshift estimate, respectively. To examine the light distribution, we chose an inclusive cluster-member sample based on $|z_{\rm b}-z_{\rm cl}|<0.1,\, z_{\rm b,min}>z_{\rm cl}-0.2,\,z_{\rm b,max}<z_{\rm cl}+0.2,\,
17<m_{F625W}<24.5$.
In order to do a simultaneous analysis of the WL signal in both the {\it HST} and Subaru fields we need to account for the different redshift distribution of the different populations these two datasets target. We do this by estimating the depth factor, $\beta(z)$, from the {\it HST} photo-z's. For the {\it HST} background catalog we estimate $\beta\sim0.525$, about a $\sim10\%$ increase in depth relative to the Subaru catalog, estimated at  $\beta\sim0.48$. We apply this relative correction to the {\it HST} catalog and scale it to match the Subaru catalog.

\begin{figure}
 \begin{center}
 \includegraphics[width=0.47\textwidth,clip]{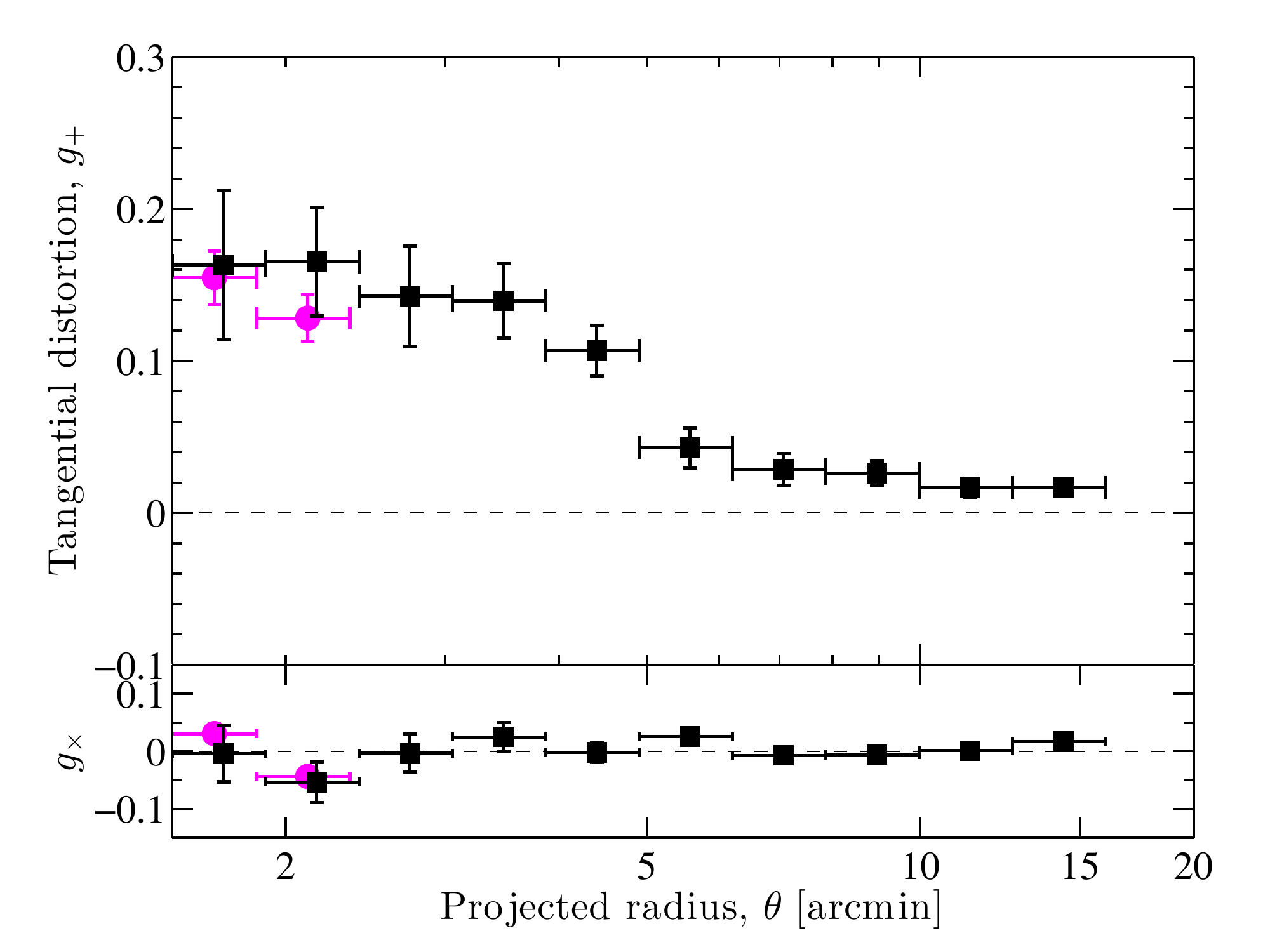}
\end{center}
\caption{As in Figure~\ref{fig:gt1D}, we show the tangential distortion profile
of the Subaru background sample (black squares), and compare with the
{\it HST}-derived background sample in the central $R<2.4\arcmin$ (magenta circles)
but outside the critical lensing region, $R>1.5\arcmin$. As can be seen, these
two complementary dataset give consistent WL signal at the region of overlap,
whereas the {\it HST} points have much smaller errors and therefore provide better
constraints.}\label{fig:gt1D_HST}
\end{figure}

A significant part of the {\it HST} region resides inside the tangential critical curve
area, and so is super-critical. In order to avoid nonlinear effects in the WL
analysis, we examine the tangential distortion profile only at the outer region
of {\it HST}, $1.5<\theta<2.5\arcmin$, to the limit of the data. We present the results in
Figure~\ref{fig:gt1D_HST}, where we overplot the inner two bins (magenta
circles) from {\it HST} that overlap with the same region in the inner Subaru (black
squares) profile. The two independent datasets show consistent WL shear signal
within the uncertainties, although we note the {\it HST} signal does show a slightly
decreased level. This may arise from nonlinear effects causing an underestimation of the shear (and possibly also the non-null cross-shear in the second bin), or simple image edge effects due to different filter coverage. However, this level is still negligible and does not affect our results significantly. We will present the 2D
distribution analysis of the {\it HST} region in section~\ref{subsec:HST2DK}, after we first
incorporate the {\it HST}+Subaru WL with {\it HST} SL information to reconstruct
the mass profile over the entire Subaru field of view (FOV) in the next section.

\section{Radial Mass Profile Analysis}
\label{sec:wl1D}

\subsection[Mass Reconstruction I]{Mass Profile Reconstruction using One-dimensional Shear and Magnification}
\label{subsec:1dk}

\begin{figure}[tb]
 \begin{center}
   \includegraphics[width=0.5\textwidth,angle=0,clip]{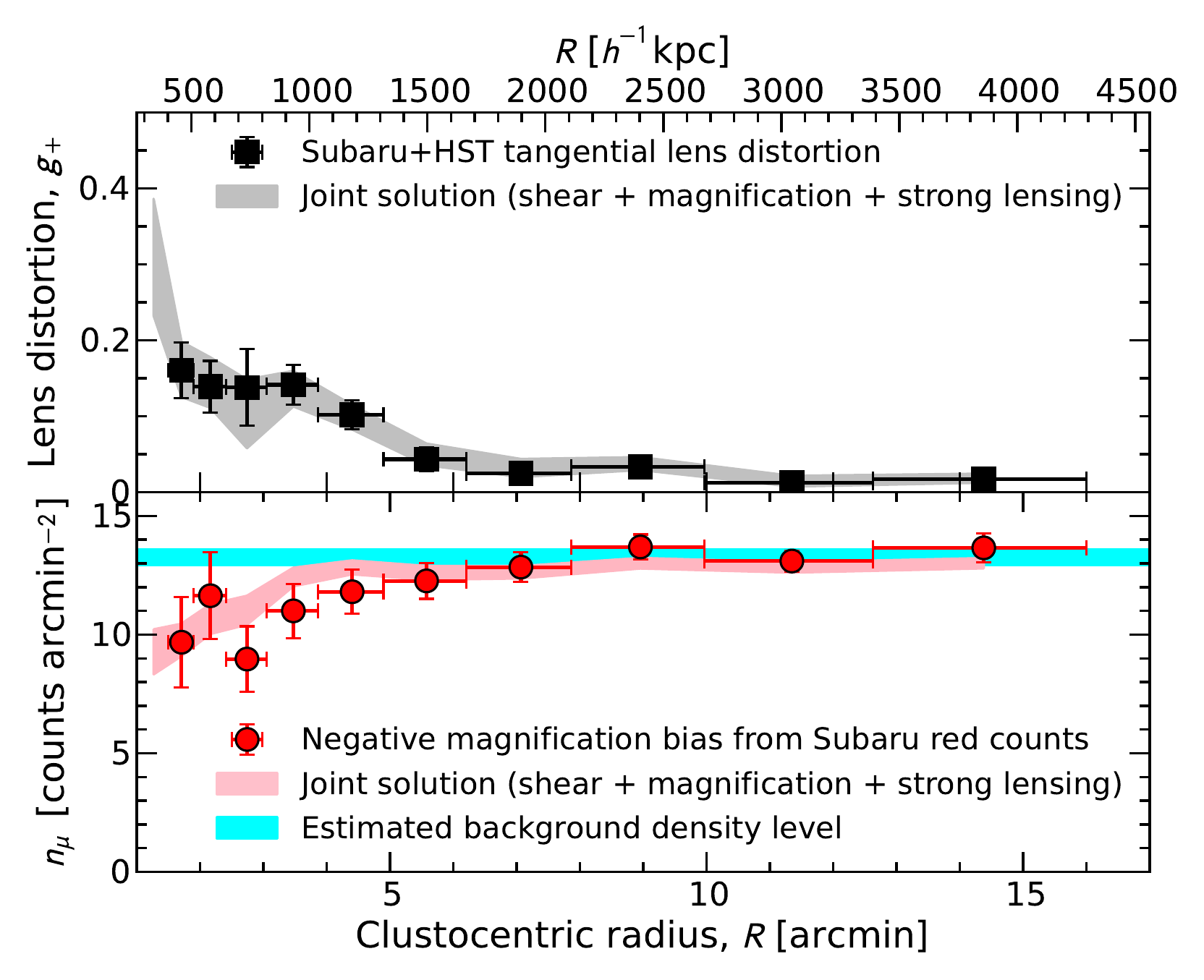}
 \end{center}
\caption{ {\it Top:}  Tangential reduced shear profile $g_+(\theta)$
(squares) based on {\it HST} and Subaru distortion data of
 the composite full background sample.
{\it Bottom:}  Coverage-corrected count profile $n(\theta)$ (circles) for a flux-limited
 sample of red background galaxies registered in the Subaru $BR_{\rm c}z'$ images.  
 The horizontal bar represents the constraints on
 the unlensed count normalization, $n_0$, as estimated from Subaru
 data.  Also shown in each panel is the joint Bayesian reconstruction (68\% CL; solid area) 
 from SL, WL tangential distortion (squares), and magnification-bias measurements (circles).}
\label{fig:wldata}
\end{figure}
\begin{figure}[tb]
\centering
\includegraphics[width=0.5\textwidth,angle=0,clip]{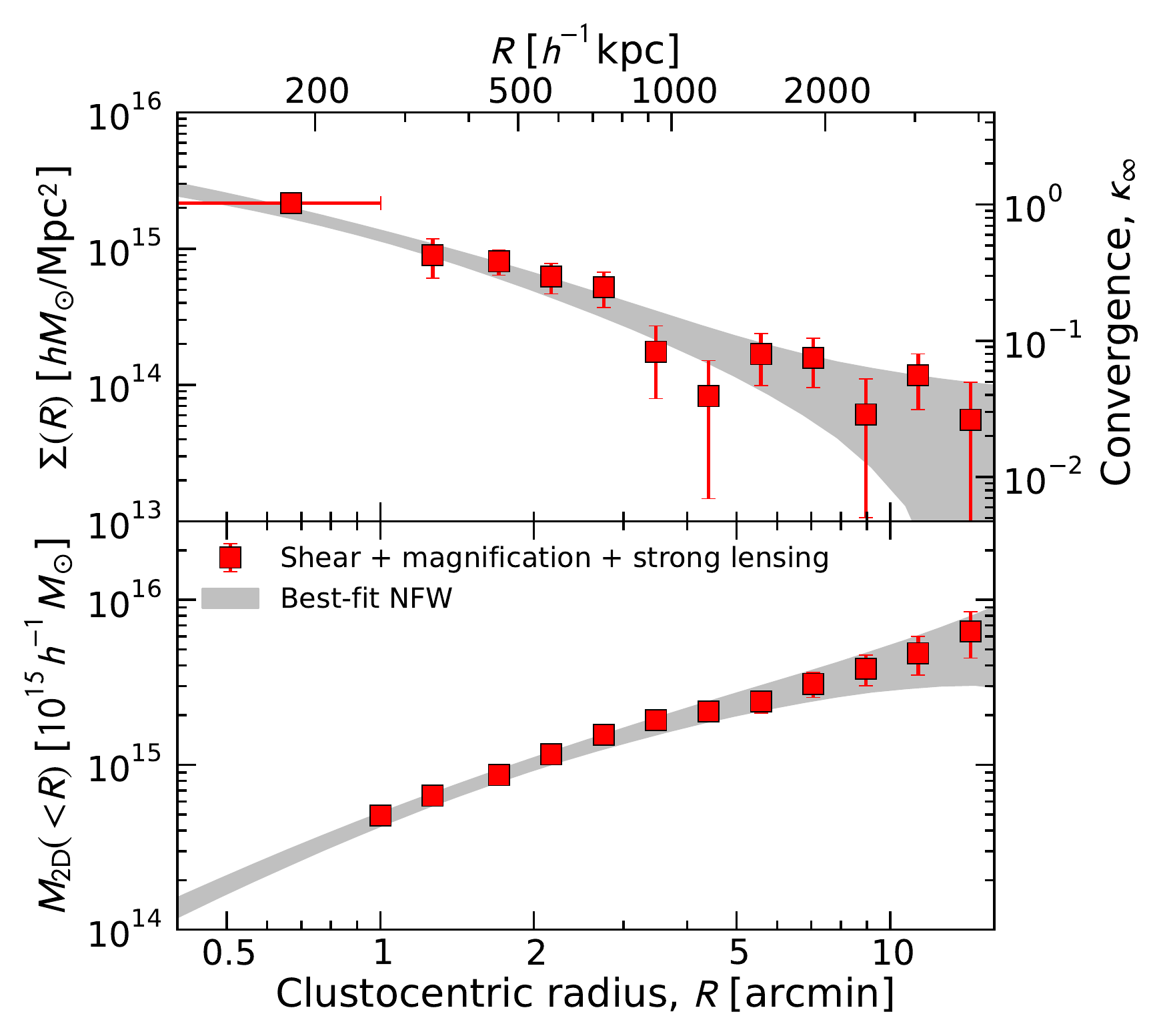}
\caption{{\it Top}: Surface mass density profile $\Sigma(R)$ (squares) derived from 
 a joint SL, WL distortion and magnification (WL+SL) likelihood analysis of our {\it HST}+Subaru lensing observations.
The gray area represents the best-fit NFW profile for the mass profile solution $\Sigma(R)$.
{\it Bottom}: The cumulative mass $M_{\rm 2D}(<R)$ (red squares) as derived from the full lensing analysis.
 The gray area is the NFW fit ($1\sigma$ confidence interval) to the WL+SL constraints as described above. }
\label{fig:kappa}

\end{figure}


We derive the cluster mass profile as a function of clustercentric radius
from a joint likelihood analysis of independent
WL distortion ($g_+$), magnification-bias ($n_\mu$), and SL projected mass ($m$) constraints,
 $\{g_{+,i}\}_{i=1}^{N_{\rm wl}}$ (Section~\ref{subsec:gt}), 
 $\{n_{\mu,i}\}_{i=1}^{N_{\rm wl}}$ (Section~\ref{subsec:magbias}),
 and
 $\{m_i\}_{i=1}^{N_{\rm sl}}$ (Section~\ref{sec:sl}),
 respectively,
following the Bayesian approach of \cite{Umetsu2013}, 
who extended the {\it shear-and-magnification} analysis method 
of \citet{Umetsu+2011} to include the inner SL information.
Such a multi-probe approach is critical for improving
the accuracy and precision of the cluster lens reconstruction, 
effectively breaking the mass-sheet degeneracy 
\citep[see ][]{2001PhR...340..291B,Umetsu+2012}.
Adding SL information to WL is
needed to provide tighter constraints on the inner density profile
\citep[e.g.,][]{UB2008,Umetsu+2011,Umetsu+2012}.
The shear+magnification method of \citet{Umetsu+2011}
has been extensively used to reconstruct the projected mass
profile in a dozen clusters \citep{Umetsu+2011,Zitrin+2011_A383,Coe2012,Umetsu+2012,Zitrin2013}.
In all cases, we find a good agreement between independent WL 
and SL mass profiles in the region of overlap.

Briefly summarizing, the model is described by a vector $\bs$ of 
parameters containing the discrete convergence profile 
$\{\kappa_{\infty,i}\}_{i=1}^N$, given by 
$N=N_{\rm wl}+N_{\rm sl}$ binned $\kappa$ values,
and the average convergence enclosed by the innermost aperture radius
$\theta_{\rm min}$ for SL mass estimates,
$\overline{\kappa}_{\infty,{\rm min}}\equiv\overline{\kappa}_\infty(<\theta_{\rm min})$, 
where we have introduced the convergence 
for a fiducial source in the far background of the cluster,
$\kappa_{\infty,i}\equiv \kappa(\theta_i;z_s\to \infty)$.
The model $\bs =\{\overline{\kappa}_{\infty,{\rm min}},\kappa_{\infty,i}\}_{i=1}^{N}$
is then specified by a total of $(N+1)$ parameters.  
Additionally we account for the uncertainty in the calibration
parameters, $\bc=(w_g,w_\mu,n_0,s)$, namely
the population-averaged lensing strengths for the distortion and magnification measurements
(see Table~\ref{tab:wlsamples}),
$w_g\equiv \langle\beta({\rm back})\rangle/\beta(z_s\to\infty)$
and
$w_\mu\equiv \langle\beta({\rm red})\rangle/\beta(z_s\to\infty)$,
the normalization and
slope parameters $(n_0, s)$ of the red-background counts (see Section~\ref{subsec:magbias}).
The covariance matrix $C_{ij}$ for the profile reconstruction is also constructed,
and used for calculating the likelihood function of the combined WL+SL observations.

In the present analysis, 
we calculate the $g_+$ and $n_\mu$ profiles
in $N_{\rm wl}=10$ clustercentric radial bins, spanning
the range $\theta=[1.5\arcmin,16\arcmin]$, 
with a constant logarithmic radial spacing $\Delta\ln\theta\simeq 0.237$.  
Additionally, 
we use our projected mass measurement within a radius of $60\arcsec$,
$m=M_{\rm 2D}(<60\arcsec)=(4.87\pm 0.35)\times 10^{14}M_\odot\,h^{-1}$
($N_{\rm sl}=1$),
tightly constrained by our detailed strong-lens modeling (Section~\ref{sec:sl}).
Note that, enclosed masses at the location around the Einstein radius
($\theta_{\rm Ein}\approx 60\arcsec$ at $z_s=2.963$ here)
are less sensitive to modeling assumptions and approaches \citep[see][]{Umetsu+2012},
serving as a fundamental observable quantity in the
SL regime \citep{Coe+2010}.
Hence, we have a total of $N_{\rm tot}=2N_{\rm wl}+N_{\rm sl}=21$ 
constraints.  The mass profile model is described by $N+1=12$ profile 
parameters and additional 4 calibration parameters ($\bc$) to marginalize over.

The resulting mass profile
$\bs$ from a joint SL+WL likelihood analysis of our {\it HST}+Subaru observations
is shown in the top panel Figure~\ref{fig:kappa} (red squares; see also Figure~\ref{fig:1d-2d}, red squares).
We find a consistent mass profile solution $\bs$ 
as displayed in Figure~\ref{fig:wldata} (solid areas). 
The projected cumulative mass profile $M_{\rm 2D}(<\theta)$  
is shown in the lower panel (red curve).
It is given by integrating the density profile
$\bs=\{\overline{\kappa}_{\infty,{\rm min}},\kappa_{\infty,i}\}_{i=1}^{N}$
\citep[see Appendices A and B of][]{Umetsu+2011} as
\begin{equation}
\begin{split}
M_{\rm 2D}(<\theta_i)&=\\
\pi (D_l\theta_{\rm min})^2& \Sigma_{\rm
 crit}\overline{\kappa}_{\rm min}
+2\pi D_l^2 \Sigma_{\rm crit}
\int_{\theta_{\rm min}}^{\theta_i}\! d\ln\theta\,\theta^2\kappa(\theta).
\end{split}
\end{equation}
The total projected mass enclosed within a radius of $\theta\approx7\arcmin\approx1.88\,\mpch$ 
is found to be $M_{\rm 2D} = (3.1\pm 0.5)\times\mhunit$.

As is evident from the cored density profile in the central region derived by SL, as well as from the flattened outer profile at large radii seen by WL (Figure~\ref{fig:1d-2d}), MACSJ0717 is a complex, non-relaxed cluster whose matter distribution is not well described by a single NFW profile. However, in order to derive a total (spherical) mass estimate of the main cluster component in a complementary (yet model-dependent) approach, we choose here to fit 
our full-lensing mass profile with a spherical NFW model.

To that end, the projected radial mass profile $\bs$ 
is fitted with a model consisting of 
a halo component described by the two-parameter universal NFW profile, $\kappa_{\rm NFW}(\theta)$,
and a constant mass-sheet component, $\kappa_c$:
\begin{equation}
\label{eq:kmodel}
\hat{\kappa}(\theta) = \kappa_{\rm NFW}(\theta) + \kappa_c,
\end{equation} 
where the constant $\kappa_c$ approximates the inherent "two-halo" term contribution due to the
clustering of halos \citep[see][]{Oguri+Hamana2011}. 
The NFW mass density profile is given by the form
\begin{equation}\label{eq:nfw}
\rho_{\rm NFW}(r)=\frac{\rho_{s}}{(r/r_s)(1+r/r_s)^{2}},
\end{equation} 
where $\rho_{s}$ is the characteristic density, and
$r_s$ is the characteristic scale radius at which the logarithmic density slope is isothermal.
The halo virial mass is then given by integrating the NFW profile (Eq.~\ref{eq:nfw})
out to the virial radius $r_{\rm vir}$,
$\mvir \equiv M(<r_{\rm vir})$.
We specify the projected NFW model with the halo virial mass, $M_{\rm vir}$, and
the degree of concentration, $\cvir\equiv r_{\rm vir}/r_s$.
We refer all our virial quantities
to an overdensity of  $\Delta_c\equiv\Delta_{\rm vir}\approx 138$ 
based on the spherical collapse model using the fitting formula by \citet[][their Appendix A]{Kitayama+Suto1996}.\footnote{$\Delta_{\rm
 vir}\approx 140$ using the fitting formula by \citet{Bryan+Norman1998}.}


We constrain the model parameters $\bp=(M_{\rm vir}, c_{\rm vir}, \kappa_c)$
with our full-lensing mass profile $\bs$.
%
The $\chi^2$ function for our SL+WL observations is\footnote{The calibration uncertainties in observational parameters,
such as the background mean depths, $\langle\beta({\rm back})\rangle$ (or $z_{s,{\rm eff}}$) and $\langle\beta({\rm red})\rangle$,
have already been marginalized over in the Bayesian mass profile reconstruction.}
\begin{equation}
\label{eq:chi2}
\chi^2(\bp) = \sum_{i,j} 
[\bs_i-\hat{\bs}_i(\bp)] 
{\cal C}^{-1}_{ij} 
[\bs_j-\hat{\bs}_j(\bp)],
\end{equation}  
where $\hat{\bs}_i$ is the model prediction
for the convergence profile $\bs_i$ at $\theta_i$,
${\cal C}$ is the full covariance matrix of $\bs$ defined as ${\cal C}=C + C_{\rm lss}$
with $C_{\rm lss}$ being the cosmic covariance matrix
responsible for the uncorrelated LSS projected along the line of sight.
For details, see \citet{Umetsu+2011_stack}.

Our best-fit NFW model to the combined lensing constraints (Equation \ref{eq:chi2})
is shown in Figure~\ref{fig:kappa} as the gray shaded area. 
To summarize our results from this analysis, 
we obtain a total virial mass estimate of  $\mvir=(2.13^{+0.49}_{-0.44})\times\mhunit \sim (3\pm 0.6)\times10^{15}M_\odot$
with the minimized $\chi^2$ of 10 for 9\,DOF
(see a summary in Table~\ref{tab:nfw}). 
We will compare and discuss all the mass estimates yielded by the different modeling and reconstruction methods 
explored in the paper in Section~\ref{subsec:1d2dcomp}.

At larger radii, a flattening of the mass profile is observed, possibly indicative of the surrounding LSS.
The deviation of the profile from a single spherical NFW halo at large radius is indicative of substructure associated with this cluster region, and therefore merits a more careful two-dimensional (2D) analysis, which we present in the next sections.

\subsection[mass reconstruction II]{Mass Profile Reconstruction using Two-dimensional Shear and Magnification}
\label{subsec:1dk2dg}

We follow \citet[][]{Umetsu+2012} to
extend the one-dimensional Bayesian method above (Section \ref{subsec:1dk})
into a 2D mass distribution by combining the 
spatial shear pattern $(g_1(\btheta),g_2(\btheta))$
with the azimuthally-averaged magnification measurements $n_\mu(\theta)$
(Section \ref{subsec:magbias}),
imposing a set of azimuthally-integrated constraints on the underlying $\kappa(\btheta)$ field.\footnote{Since the degree
of magnification is locally related to $\kappa$, this will essentially
provide the otherwise unconstrained normalization of $\kappa(\btheta)$
over a set of concentric annuli where count measurements
are available. We note that no assumption is made of azimuthal
symmetry or isotropy of the cluster mass distribution.}
For details of the method, we refer the reader to 
\citet[][their Appendix A.2]{Umetsu+2012}.

By combining complementary WL distortion and magnification data in a non-parametric manner, 
we construct a 2D mass map over a $48\times 48$ grid 
with $0.5\arcmin$ spacing
covering the central $24\arcmin\times 24\arcmin$ field.\footnote{The magnification analysis is 
limited within the central $24\arcmin\times 24\arcmin$ region where
the number counts of red background galaxies are reliably measured.}
We show in Figure~\ref{fig:1d-2d} (black circles) the azimuthally-averaged radial mass profile 
$\Sigma(R)$ produced from the resulting mass map, 
given in 10 linearly-spaced radial bins spanning
from $\theta=1.5\arcmin$ to $16\arcmin$.  The innermost bin with the
horizontal bar represents the mean interior mass density $\overline{\Sigma}(<1.5\arcmin)$
as marked at the area-weighted center $\overline{\theta}=1\arcmin$.  
We find a model-independent constraint on the total enclosed mass within $\theta\approx7\arcmin$ 
to be $M_{\rm 2D} = (3.4\pm0.6)\times\mhunit$. 

We fit an NFW + mass-sheet model (Equation~\ref{eq:kmodel})
to the WL radial mass profile derived here 
by minimizing the total $\chi^2$ function defined as in Equation~(\ref{eq:chi2}),
marginalizing over the mean background depth uncertainty in $z_{s,{\rm eff}}$.
We find the total virial mass
to be $\mvir=(2.23^{+0.44}_{-0.38})\times\mhunit$.
We summarize all the results from our analyses in Table~\ref{tab:nfw} and discuss the differences in Section~\ref{subsec:1d2dcomp}.


\section{Spatial Mass Distribution Analysis}
\label{sec:wl2D}

Here we present the full 2D mass distribution in the cluster and its
surrounding structures as constrained by WL shear in \ref{subsec:2DK},
and we further constrain the mass of the individual components by
modeling identified halos independently in \ref{subsec:2Dhalo}.  We
present another 2D inversion method, SawLenS, in \ref{subsec:sawlens}
which incorporates information from WL shear and SL on a multi-scale grid. Finally we probe the substructure in the core of the cluster in \ref{subsec:HST2DK}.

\subsection{Mass and Light Distributions of the Cluster and Surrounding Large-Scale Structure}
\label{subsec:2DK}
\begin{figure*}[htb]
\centering
\includegraphics[width=0.5\textwidth]{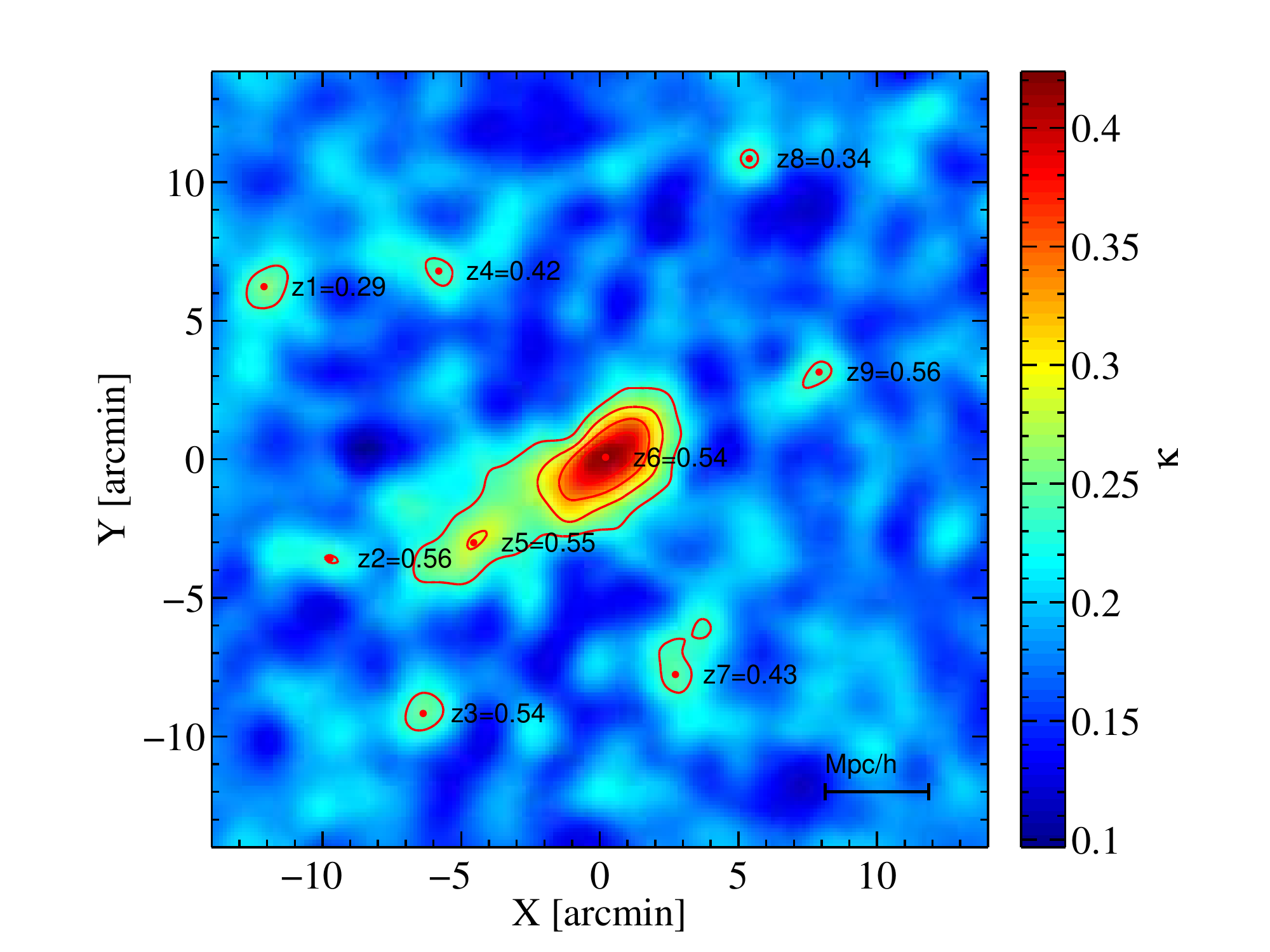}
\hspace{-0.5cm}
\includegraphics[width=0.5\textwidth]{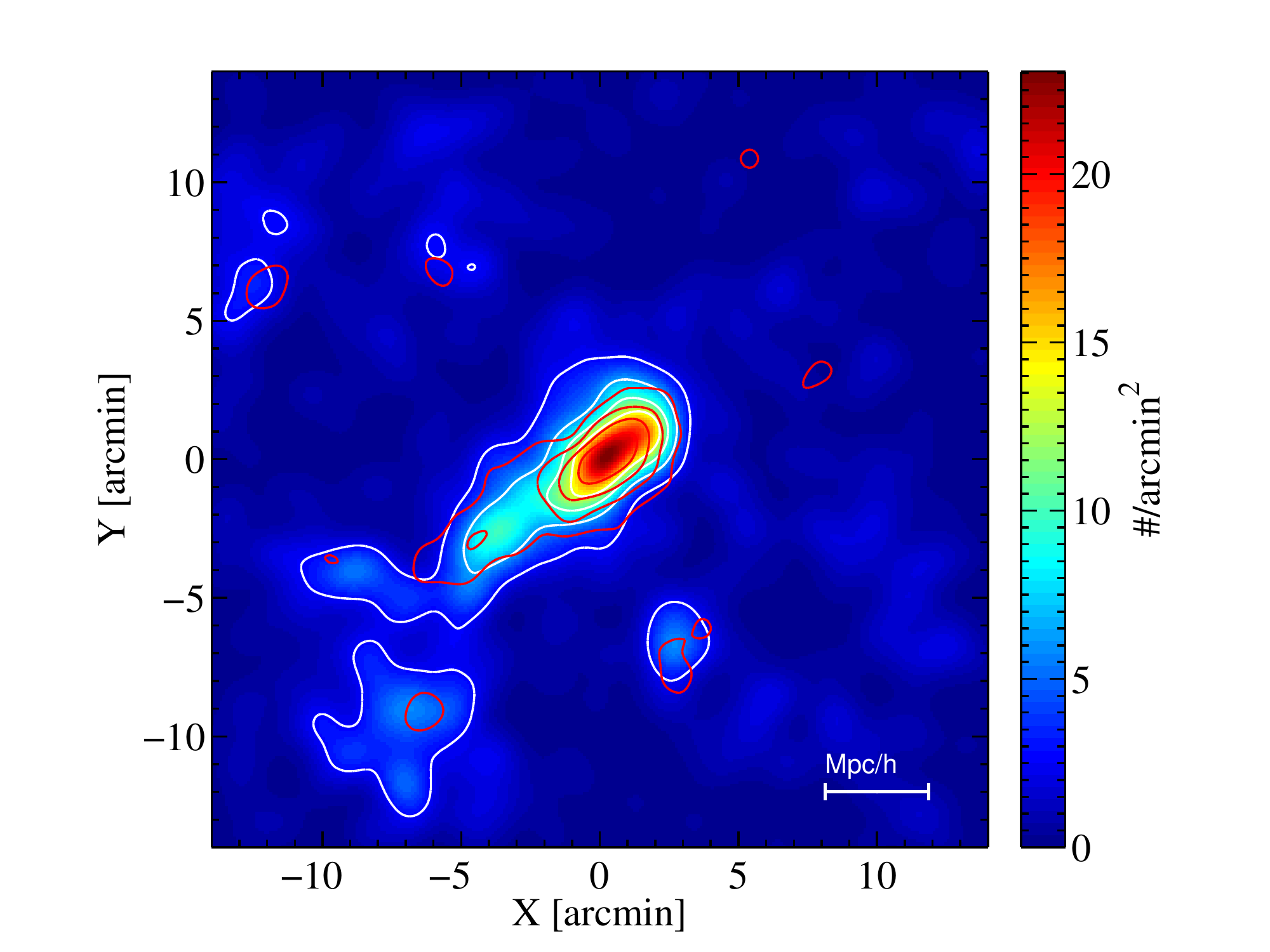}
\includegraphics[width=0.5\textwidth]{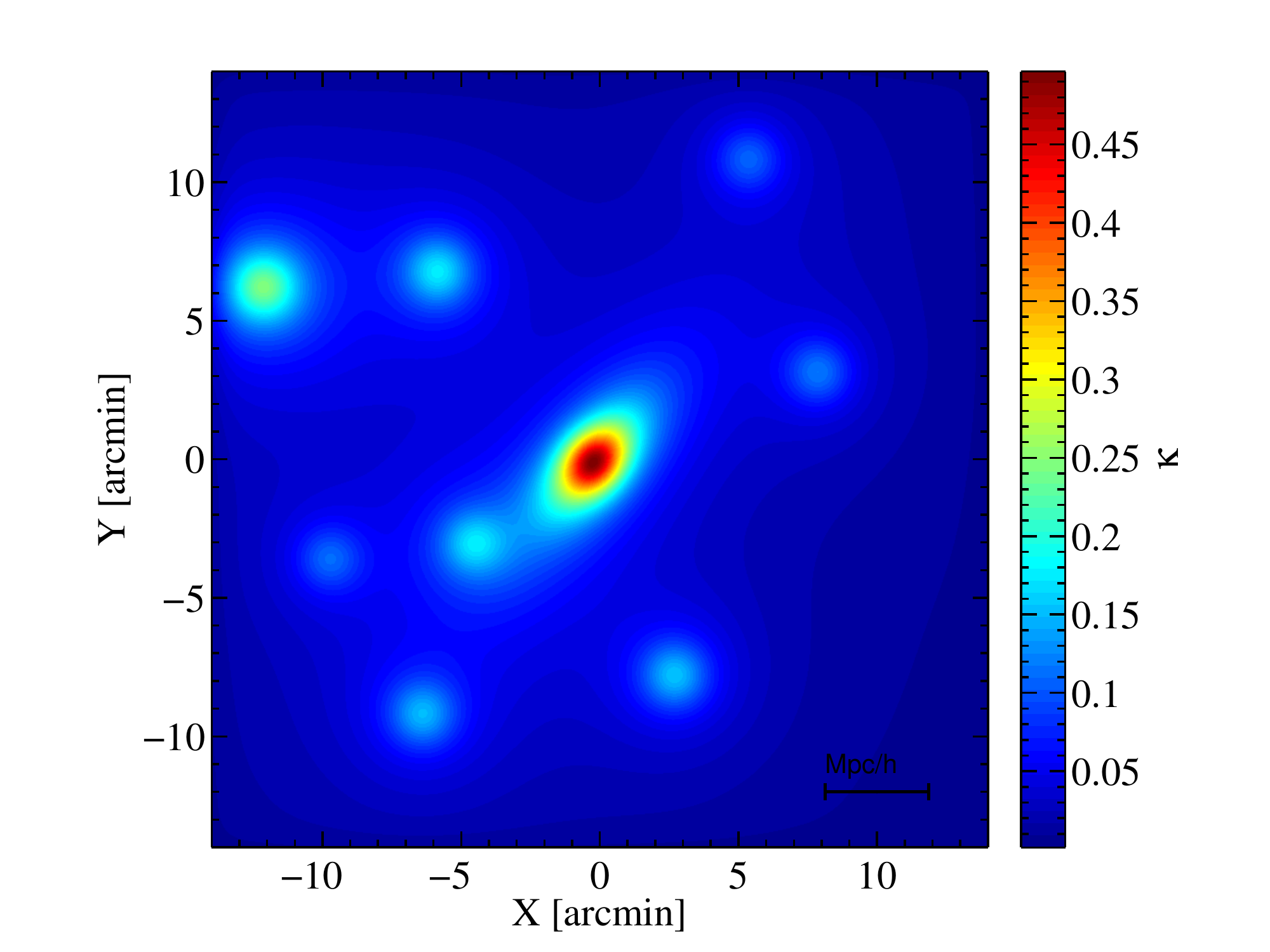}
\hspace{-0.5cm}
\includegraphics[width=0.5\textwidth]{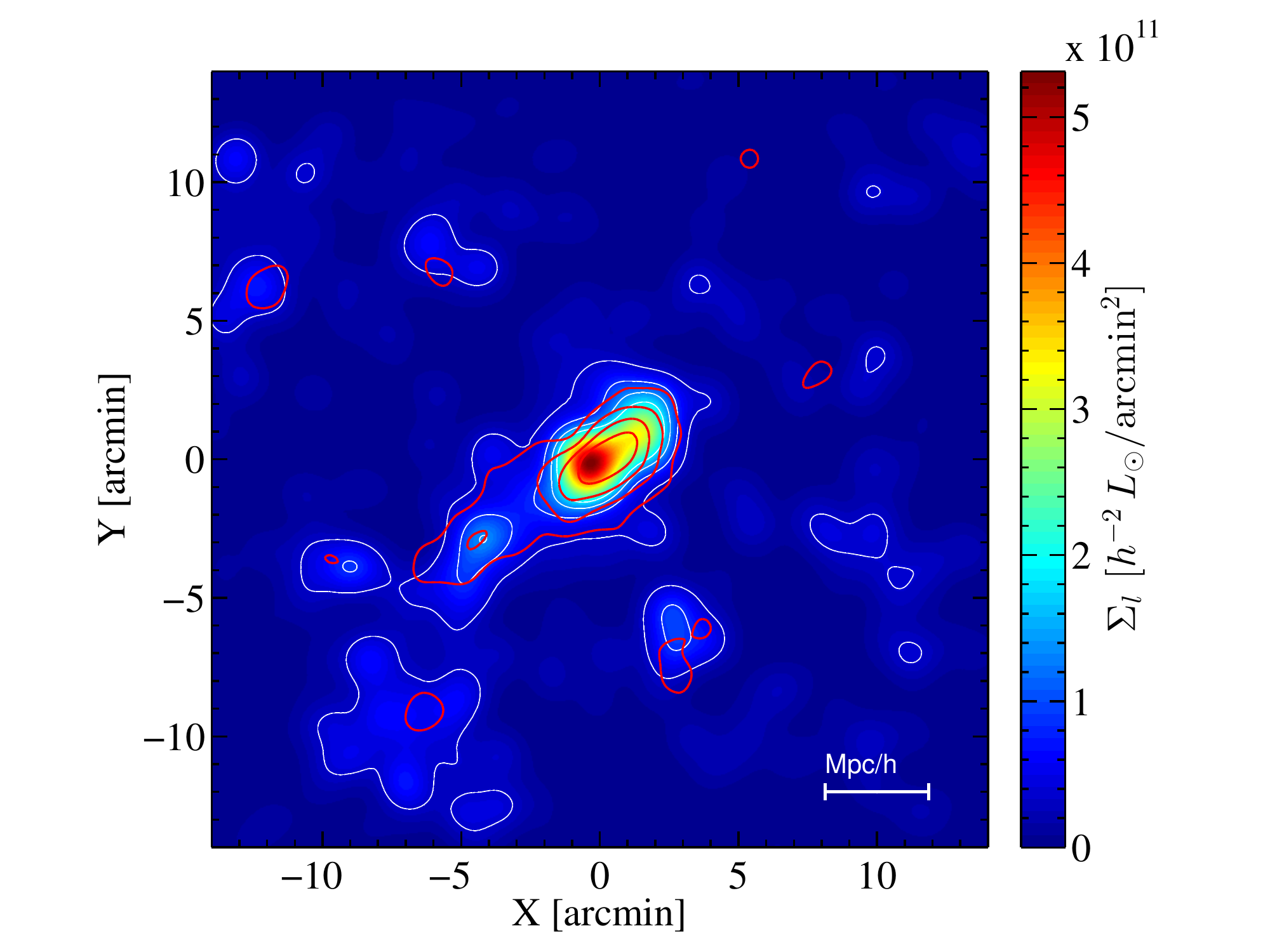}
\caption{{\it Top Left}: linear reconstruction of the dimensionless surface-mass density
distribution, or the lensing convergence $\kappa(\btheta)=\Sigma(\btheta)/\Sigma_{\rm crit}$, reconstructed from Subaru distortion data. The lowest red contour represents our detection level at $2.5\sigma_\kappa$, with further increments of $2\sigma_\kappa$. Red points denote the locations of the 9 significant mass peaks identified above $2.5\sigma_\kappa$. 
{\it Top Right}: Surface number density distribution $\Sigma_n(\btheta)$ of green galaxies, representing cluster member galaxies. White contours show $5\sigma_n$ increments starting at $2.5\sigma_n$. Also overlaid are the red contours of the convergence, showing very good agreement between the DM and the galaxy distributions.
{\it Bottom Left}:  Dimensionless surface-mass density distribution reconstructed from the multi-halo shear-fitting analysis, comprising the NFW halos fitted to the 9 main mass peaks detected in \ref{fig:kappa2D}. The most massive central component was fitted by an elliptical-NFW model, whereas the other peaks are fitted by a simple NFW model. The parameters of each halo fit are given in Table~\ref{tab:nfw}.
{\it Bottom Right}: Luminosity density distribution $\Sigma_l(\btheta)$ of green galaxies, representing cluster member galaxies. White contours show $5\sigma_l$ increments starting $2.5\sigma_l$ in number density. Also overlaid are the red contours of the convergence.
}
\label{fig:kappa2D}
\end{figure*}

WL distortion measurements ($g$) can be used
to recover the underlying projected mass density field
$\Sigma(\btheta)$.
Here we use the linear map-making method outlined in Section 4.4 of
\citet[][see their Section 4.4]{Umetsu+2009} to
derive the projected mass distribution from
the {\it HST}+Subaru distortion data presented in the previous sections.

In Figure~\ref{fig:kappa2D} (top left panel),
we show the surface-density field $\kappa(\btheta)$ in the central $28\arcmin\times
28\arcmin$ region,
reconstructed from the background (blue+red) sample (Section~\ref{subsec:color}),
where for visualization purposes the mass map is smoothed with a
Gaussian of $\theta_{\rm FWHM}=1.5\arcmin$. We overlay the mass map with contours starting at $2.5\sigma_\kappa$ above the background, which is equivalent to a detection threshold of $\Sigma_{2.5\sigma_\kappa}=3.66\times10^{14}\,h\,M_\odot/{\rm Mpc}^2$, and separated by $2\sigma_\kappa$ intervals.

A very elongated structure is evident, and several mass clumps seem to comprise the cluster.  Some clumps lie outside of the estimated virial radius which is at $\sim2\,$Mpc/$h$. Some of these mass structures may in fact lie in the foreground of the cluster and not be physically associated with it, yet contribute to the overall lensing of the background galaxies. We therefore need to estimate the mean redshift of each of these significant structures.

In order to estimate the distances to the detected structures, and to correlate the galaxy distribution in this cluster and its surrounding structure with the DM distribution, we utilize the green sample defined in Section~\ref{subsec:clsample}, which comprises mostly cluster galaxies and some lower-level contamination of foreground galaxies. We construct both a 2D galaxy number-density map (Figure~\ref{fig:kappa2D}, top right panel) and a $K$-corrected $\RC$-band luminosity density map (Figure~\ref{fig:kappa2D}, bottom right panel). Both maps are smoothed with a Gaussian of the same scale as the mass map above, $\theta_{\rm FWHM}=1.5\arcmin$ . White contours are overlaid with $5\sigma$ increments starting at $2.5\sigma$ above the background level of the equivalent map. We also overlay the surface mass density map (red contours) as determined above to illustrate the correlation.

Overall, the DM distribution is well traced by the galaxy distribution, as evident by the overlapping contours and the proximity of the mass peaks to the light peaks. The good agreement seen is consistent with the smaller mass halos being in the process of accreting onto the main halo.
A further detailed analysis of individual structures is presented below (Section~\ref{subsec:2Dhalo}).


\subsection{Multi-Halo Mass Reconstruction}
\label{subsec:2Dhalo}

To estimate individual masses of the structures comprising the cluster and its surroundings as seen within the scope of the Subaru, $\sim0.25$\,deg$^2$, we now present a multi-halo fitting approach.
We first identify the most dominant peaks from the convergence map presented above (Section~\ref{subsec:2DK}, Figure~\ref{fig:kappa2D}). These are defined as all peaks lying $2.5\sigma_\kappa$ above the background level, estimated from a bi-weighted scale and mean \citep{1990AJ....100...32B}, respectively, outside the region of $8\arcmin$ from the cluster center, so as not to be biased by the cluster potential. The $2.5\sigma_\kappa$ detection level is indicated as the first red contour in Figure~\ref{fig:kappa2D}, corresponding to $\Sigma_{2.5\sigma} = 3.66\times10^{14}\,h\,M_\sun/{\rm Mpc}^2$.

We have detected nine distinct mass peaks, numbered $z_1$--$z_9$
in Figure~\ref{fig:kappa2D} (top left, the peak locations are marked with red points) and in Table~\ref{tab:nfw}. 
To help identify if these peaks are part of the same structure as the cluster or unassociated systems
projected along the line of sight, 
we estimate the photometric redshift of each component. 
For this we take the mean photometric redshift of matching member galaxies 
in the green sample (section~\ref{subsec:clsample}) lying within $1\arcmin$ from the respective mass halo peak. 
As evident, only five of the nine structures lie at approximately the same redshift $(z_2,z_3,z_5,z_6,z_9)$ as that of the cluster, 
whereas the other mass clumps lie at $z_1=0.29,z_4=0.42,z_7=0.43,z_8=0.34$, in the foreground of the cluster.

We then preform a 2D shear analysis to constrain the mass properties of the cluster and its surrounding structure,
modeled as the sum of 9 mass halos in projection space.
The multi-halo shear modeling procedure described here is similar to those of
\citet{Watanabe+2011}, \citet{Okabe+2011_A2163}, and \citet{Zitrin+2012_CCL},
but including an elliptical halo model as described below.
More details will be presented in our forth-coming
paper (Umetsu et al. 2013, in preparation). 

First we construct a reduced-shear map $(g_1(\btheta),g_2(\btheta))$ 
on a regular grid of $N_{\rm cell}=42\times 42$ independent cells,
covering a $28\arcmin\times 28\arcmin$ region centered on the cluster.
We exclude from our analysis those cells lying within $1.5\arcmin$ from the
cluster center and those lying within $0.5\arcmin$ from the other less-massive halo peaks,
to avoid potential systematic errors due to contamination by unlensed cluster member galaxies
(Section \ref{sec:samples}).
This leaves us with a total of 1813 usable measurement cells, corresponding to $3626$
constraints.

We describe the primary mass peak, responsible for the main cluster,
as an elliptical NFW (eNFW, hereafter) model
specified with 6 parameters, namely 
the halo virial mass ($M_{\rm vir}$), concentration ($c_{\rm vir}$), ellipticity ($e=1-b/a$), position angle
of the major axis ($\theta_e$), and centroid position ($X_c,Y_c$).
We introduce the mass ellipticity $e$ 
in the isodensity contours of the projected NFW profile $\Sigma(X,Y)$
as $R^2=(X-X_c)^2+(Y-Y_c)^2\to (X-X_c)^2(1-e) + (Y-Y_c)^2/(1-e)$ \citep[see][]{Oguri+2010_LoCuSS,Umetsu+2012}.
For each of the other 8 less massive halos, we assign a spherical NFW profile
parametrized with the virial mass ($M_{\rm vir}$), where the
centroid position is fixed at the respective mass peak location.
Since these less massive halos are less resolved by our WL observations,
their concentration parameters are set according to the mass-concentration relation 
$c_{\rm vir}(M_{\rm vir},z)$ given by \cite{Duffy08}.

We use the MCMC technique
with Metropolis-Hastings sampling to constrain the 
multi-halo lens model from a simultaneous 9-component fitting to the 
reduced-shear map.   The best-fit parameters are reported in Table \ref{tab:nfw}.
We find virial masses of $\mvir = (1.71\pm0.26)\times 10^{15}\,M_\sun/h$ for the main cluster halo, $z_6$,  and masses of, $\mvir=(0.15\pm0.09),(0.27\pm0.11),(0.25\pm0.11),(0.19\pm0.10)\times 10^{15}\,M_\sun/h$ for the other smaller halos, $z_2,z_3,z_5,z_9$, respectively, that lie at the same redshift.
The high value of the ellipticity , $e=0.59\pm0.08$, inferred from our elliptical lens modeling (similar to the cases of e.g., MACS J0416 by \citealt{Zitrin2013}; RX J0152.7-1357 by \citealt{Jee2005a} and \citealt{2005PASJ...57..877U}) supports that the central component of MACS0717 represents a merging, interacting system of multiple clumps in the process of formation \cite[as shown in][]{Ma2009,2012ApJ...761...47M}.

Comparing the cluster luminosity and galaxy maps, most of the significant mass structures detected in the mass map are also probed by the galaxies, with the exception of two lower-significance structures, marked as $z_8$ and $z_9$. The mass structure $z_8$ is not evident in the galaxy density map, but can be explained by the existence of lower-redshift galaxies seen therein, $z_8=0.34$, which is close to the limit probed by our color-selection method (see Figure~\ref{fig:zdist}, green). The other mass structure, $z_9$, is not detected in the galaxy density map, however, a low-significance peak is seen (more evident in the luminosity map) offset by $\sim1.5\arcmin$ to the north-east, where visual inspection reveals a bright cluster galaxy (BCG), albeit also at $z\approx0.34$, therefore not part of the cluster structure. A $\sim1.5\arcmin$ offset is comparable to our smoothing scale, therefore within errors.

\begin{figure}[tb]
\centering
\includegraphics[width=0.5\textwidth]{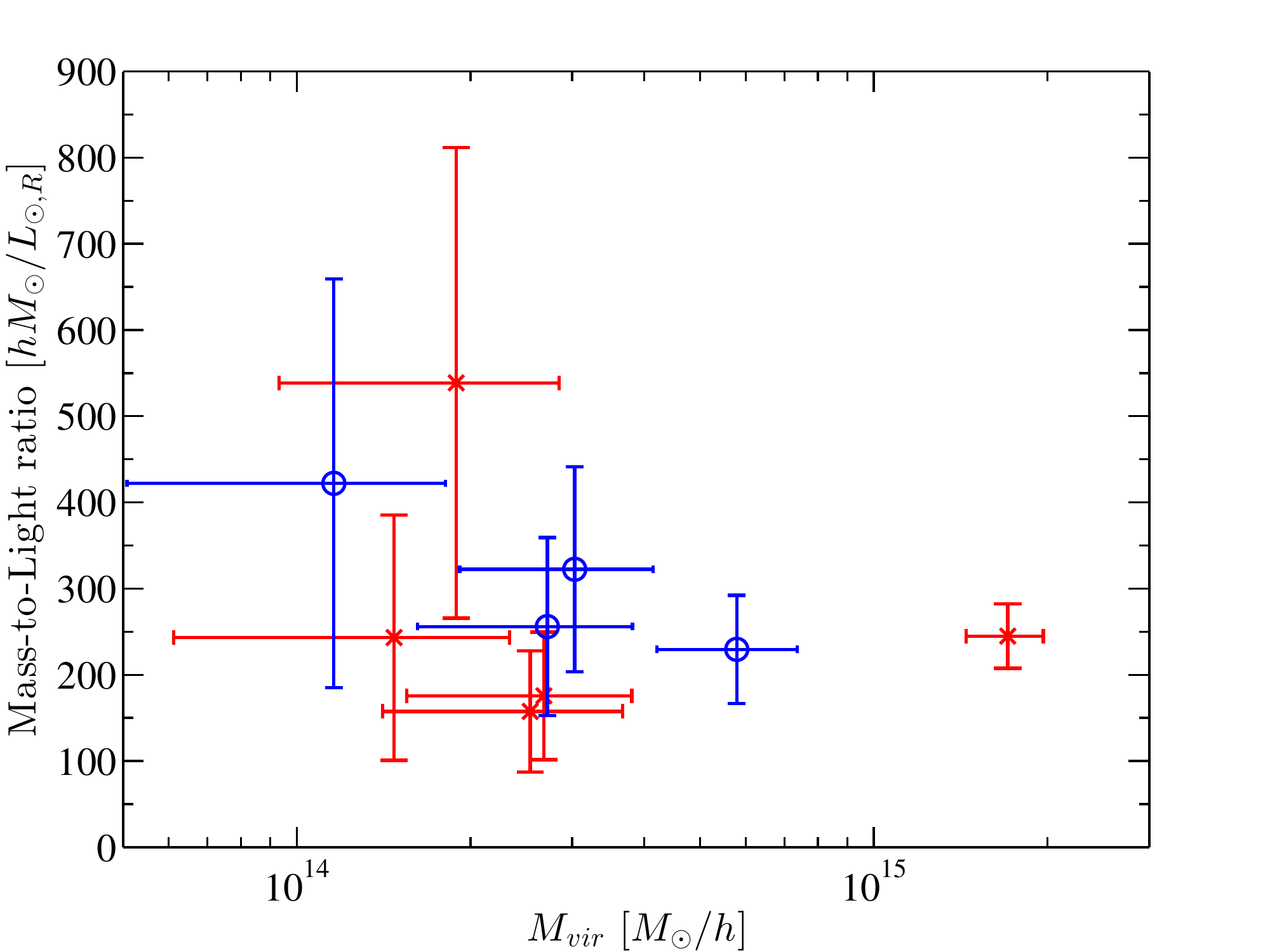}
\caption{Mass-to-light $M/L$ vs. $\mvir$ for the 9 halos modeled in Section~\ref{subsec:2Dhalo}. The mass halos that lie at the cluster redshift are shown as red stars, and the mass halos at the foreground of the cluster are plotted with blue circles. The first two points at $\sim1\times10^{14}\,M_\sun/h$ correspond to the lower-significance halos that have no optical counterpart (see text).}
\label{fig:ML_MH}
\end{figure}

To further determine if the detected and modeled halos are real, and if their content represents the typical stellar content of cluster-sized halos,
we calculate the mass-to-light ($M/L$) ratio of the individual halos modeled. We divide the fitted virial mass of each halo, $\mvir_{,i}$, by the total luminosity of ``green'' galaxies within the equivalent virial radius of the halo (virial ellipse in the case of the main halo modeled with eNFW, $z_6$), $L_{i}=\sum_{k\in r<\rvir_{,i}}{L_{\RC,k}}$. We list the $M/L$ ratios in Table~\ref{tab:nfwMH} and we plot the resulting $\mvir/L$ vs. $\mvir$ in Figure~\ref{fig:ML_MH}. All the halos, apart from halos $z_8$ and $z_9$ discussed above, show values that are expected for group- to cluster-sized halos, ranging from $\approx200-300$ \citep{Rines+2000,2004AJ....128.1078R,2004ApJ...600..657K}. However, the large uncertainties in the $M/L$ ratios measured, especially for the low-mass halos, limit us from determining if there is any trend in this relation.

For consistency check, we also calculate the $M/L$ ratios for the
single-halo NFW models we fit to our reconstructed mass profiles in
subsections~\ref{subsec:1dk} and \ref{subsec:1dk2dg} (listed in Table~\ref{tab:nfw}), and compare with
the $M/L$ ratio we get for the main halo from our multi-halo modeling, $M/L_{6}\sim245$. Within the errors, we get
comparable $M/L$ values for the one-dimensional NFW fit
(subsections~\ref{subsec:1dk}), $M/L\sim300$, which are reasonable for
a cluster as massive as MACSJ0717, whereas the values for the NFW fit to
the 2D mass reconstruction method given in
subsection~\ref{subsec:1dk2dg}, $M/L\sim310$, are only slightly higher,
reflecting the higher total mass found, due to the inclusion of
the second massive halo at larger radii in the virial mass fitted.

\subsection{SaWLens SL+WL reconstruction}
\label{subsec:sawlens}
We preform a complementary joint SL+WL reconstruction using the method of \cite[][hereafter SaWLens]{Merten+2009,Merten11}.
The SaWLens method combines central SL constraints
from multiple-image systems with WL distortion
constraints in a non-parametric manner to reconstruct
the underlying lensing potential on a multi-scale grid. Here, the density of constraints sets our resolution scale. Inside the {\it HST} region, we combine constraints from {\it HST} WL shape distortions with those of SL multiple-image sets. This yields a very high-resolution scale of $9\arcsec$ box size. Beyond this, we rely on Subaru WL shape distortion constraints, yielding a resolution scale of $25\arcsec$.
There is good consistency between the mass structures probed by the mass map derived here and in Section~\ref{subsec:2DK}. We further derive azimuthally-averaged mass profile, in order to compare with our other mass profiles derived with the methods presented in the previous sections (see Figure~\ref{fig:1d-2d}, blue triangles). 
SaWLens error bars were derived from 1000 bootstrap realizations of the WL input catalog and from 2000 re-samplings of the SL input data. 
Since many of the multiple image systems have only (less certain) photometric redshifts as derived from HST imaging, SL realizations were obtained by randomly assigning redshifts within the equivalent uncertainties of each system.
We further present the inner substructure seen by this method in the next subsection.


\subsection{Mass and Light Distribution inside the Cluster Core}
\label{subsec:HST2DK}

In this subsection, we present the 2D analysis of the WL shear map as measured
from background galaxies in the {\it HST} field, defined in section~\ref{sec:wl-hst}. We also present the inner region of the SaWLens reconstruction as an independent analysis of the cluster core.

To avoid systematic effects due to the finite-field of the {\it HST} FOV, we combine
our {\it HST} background sample with the Subaru background sample (Section~\ref{sec:wl-hst}), where we remove
Subaru objects that match with {\it HST} objects to avoid duplication. We also
correct for the different depths of the {\it HST} relative to the Subaru background
samples, as explained in Section~\ref{sec:wl-hst}.
In order to avoid infinite noise issues we first smooth the 2D shear field with
a Gaussian of $0.25\arcmin$ FWHM. This essentially limits our sensitivity
to mass structure at small angular scales, depending on the local 
source number density (${\rm S/N}\propto n_g^{1/2}$).

\begin{figure*}
 \begin{center}
 \includegraphics[width=0.5\textwidth,clip]{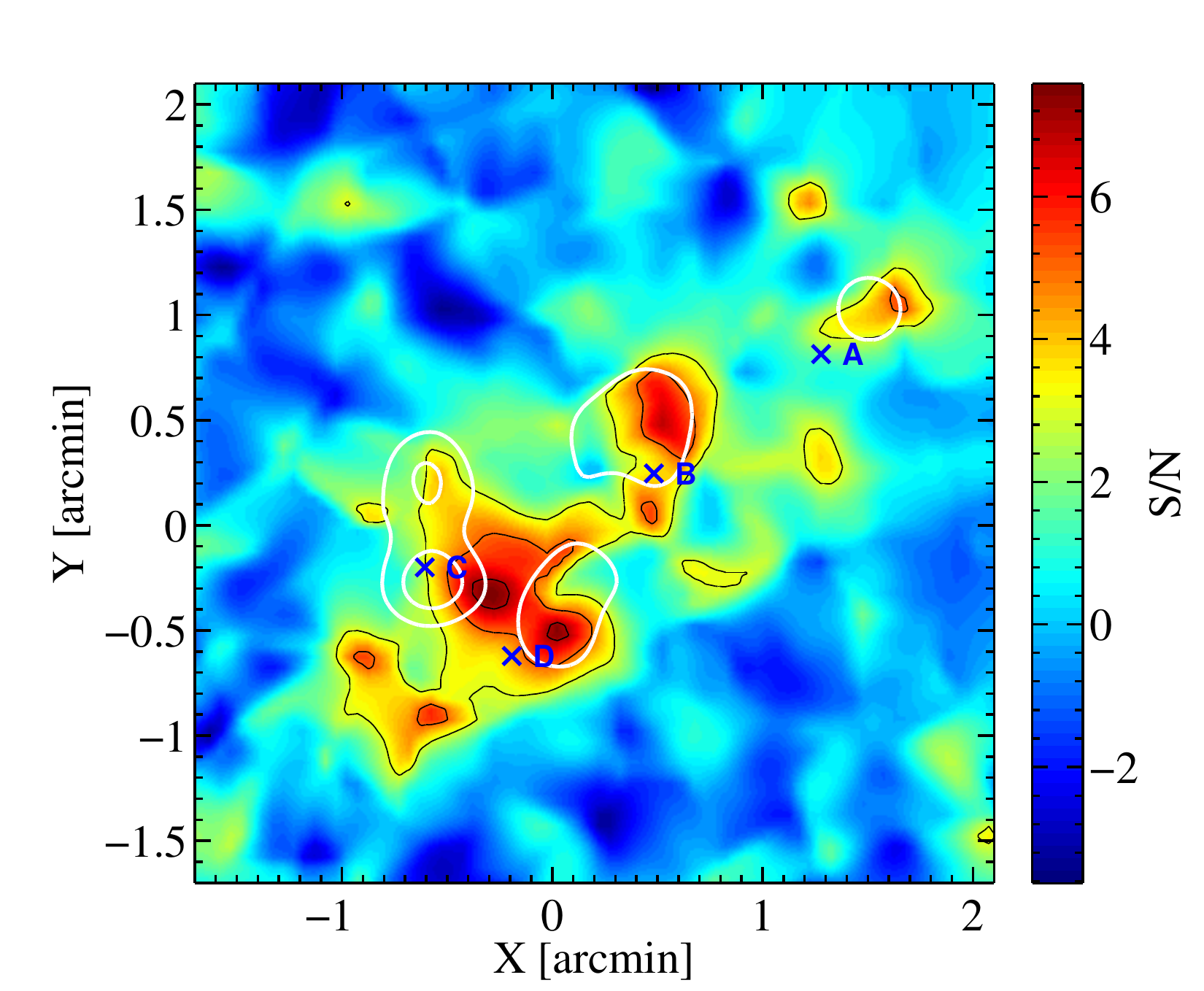}
\hspace{-0.2cm}
   \includegraphics[width=0.5\textwidth,clip]{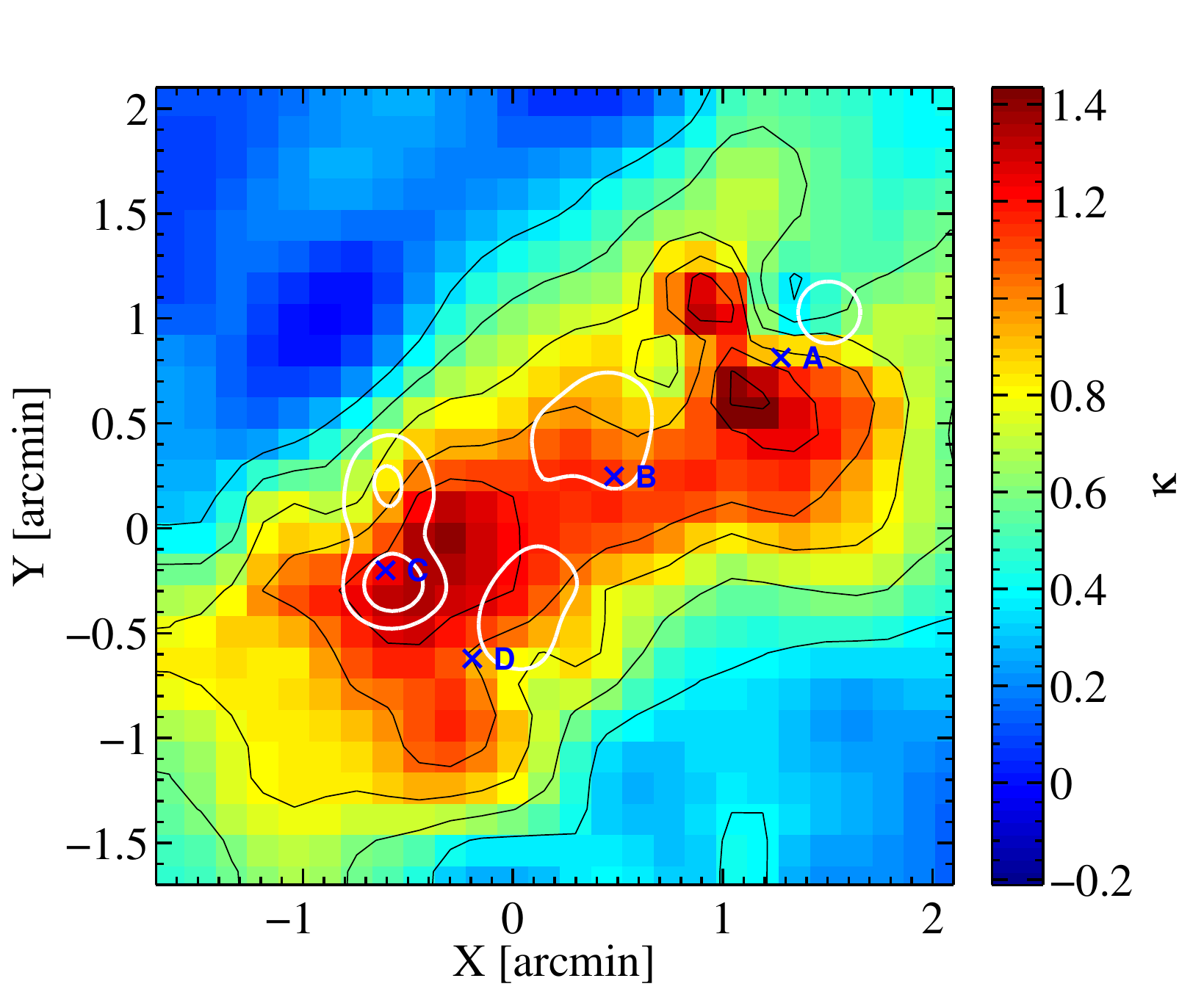}
 \includegraphics[width=0.5\textwidth,clip]{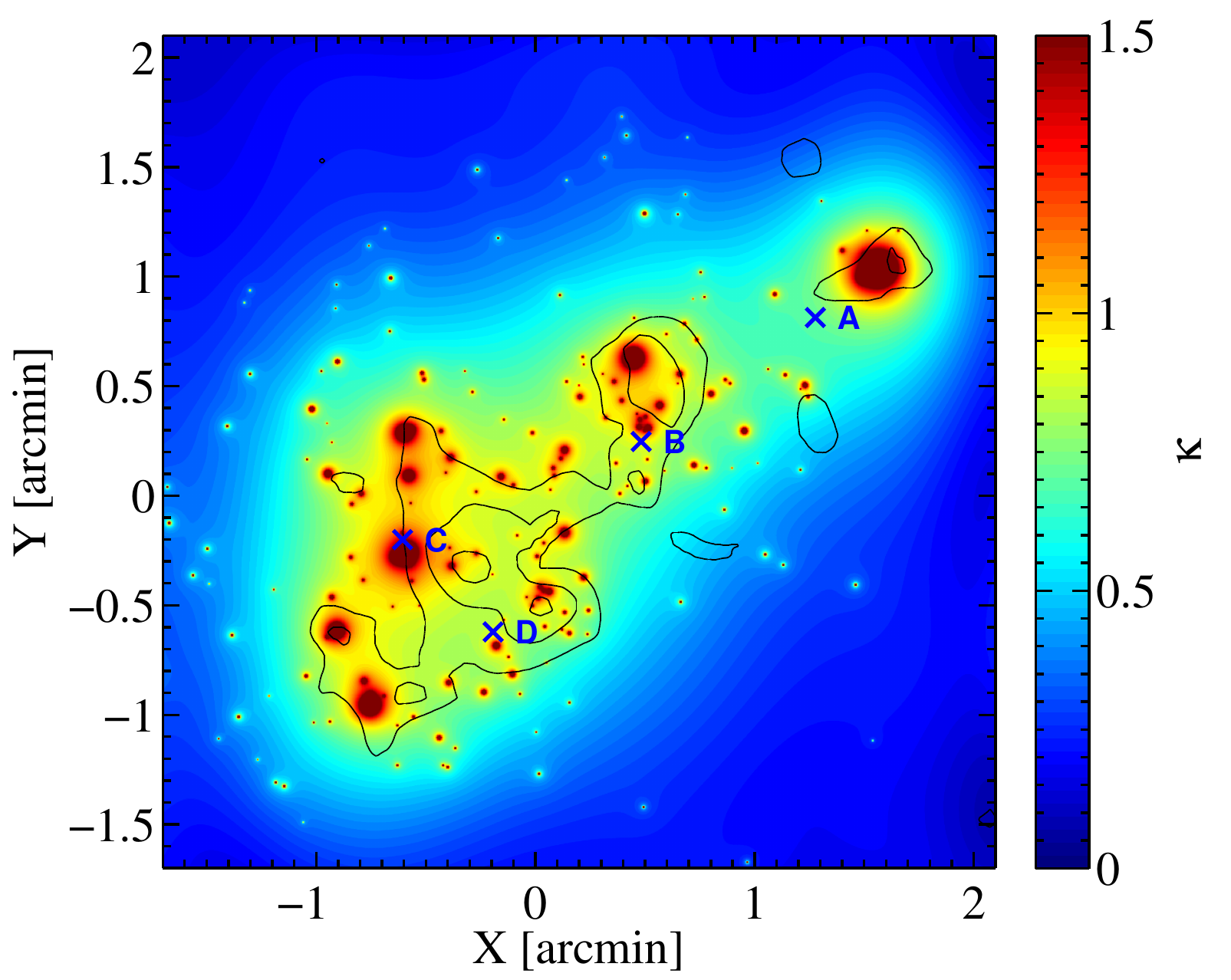}
\hspace{0.2cm}
 \includegraphics[width=0.46\textwidth,clip]{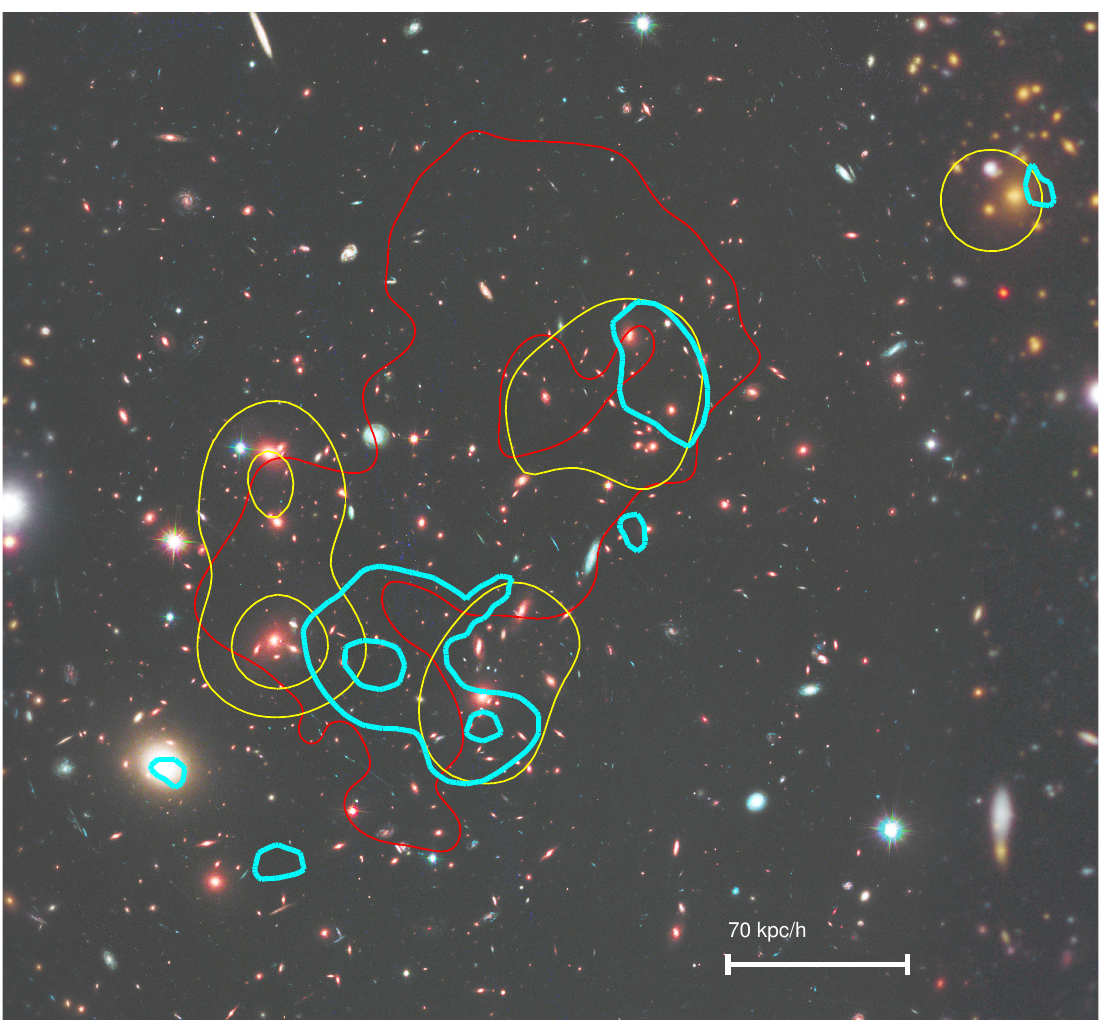}
\end{center}
\caption{{\it Top left}: S/N map of the WL MEM reconstructed mass-density distribution. The black contours  start at $3\sigma_\kappa$ with $2\sigma_\kappa$ intervals. We also plot the light contours (white) and the SL mass peak (A--D) found by L12. 
{\it Top right}: The SaWLens SL+WL reconstructed mass-density distribution, light contours (white) and L12 SL mass peak. 
{\it Bottom left}: The parametric SL model derived in Section~\ref{sec:sl}, using the light from cluster galaxies as a proxy of the mass.  The WL MEM mass reconstruction (black contours) and L12 SL mass peaks are also overlaid.
{\it Bottom right}: {\it HST} color image of MACSJ0717.
Cyan contours show the DM distribution as determined from the WL shear analysis. Yellow contours show the light distribution of cluster members. Red contours show the Chandra X-ray Luminosity map. The cluster DM component is comprised of several mass clumps, with several offsets between the DM, gas and galaxies. Notably, the central DM core lies between the light peaks.
}\label{fig:K2D_HST}
\end{figure*}

We reconstruct the 
high-resolution mass map in the central $5\arcmin\times5\arcmin$ region from 
the 2D shear map,
using an entropy-regularized maximum-likelihood reconstruction method 
(hereafter, maximum entropy method, or MEM) of \cite{UB2008}. 
We present the resulting mass
distribution in Figure~\ref{fig:K2D_HST} (upper left) in terms of the reconstruction 
S/N map, with $3,5,7\sigma_\kappa$ mass contours overlaid.  The errors for the mass map are based on the
theoretical covariance matrix \citep{UB2008}.
As an independent comparison, we also present the inner region of the mass map derived from the SaWLens method, described in the previous section. This is shown in Figure~\ref{fig:K2D_HST} (right panel).
A grid-size of $\sim9\arcsec$ is possible in this region due to several SL multiple-image systems, augmented by the {\it HST} measured WL galaxy shapes. We overlay the light distribution in both maps (white contours) as determined from the cluster members selected in section~\ref{sec:wl-hst}.

Broadly, both maps show a mass distribution that follows the light as determined by the member galaxies.  
Substantial substructure is probed by both methods, and several distinct mass peaks are evident throughout. The complexity of the mass distribution demonstrates that  a clear center is not well-defined in this cluster, which instead of a dominant BCG has several bright galaxies associated with the different mass clumps.
%
This level of substructure was previously reported by \cite{Ma2009} from Chandra X-ray
data, showing several brightness peaks (see their Figure 1) and an extended shock feature. We overlay the X-ray map (also rederived here in Section~\ref{subsec:xray} below) on the {\it HST} color image (Figure~\ref{fig:K2D_HST}, bottom right).
Combining with dynamical measurements, \cite{Ma2009} concluded the existence of 4 main
mass halos (denoted A, B, C, and D in their figure) suggesting a triple-merger. 
One of those clumps (B) was shown to still be in its
infancy at very high collision velocity of $s\sim3000$\,km$\,s^{-1}$, infalling
through the cluster.
Using their SL model, L12 also find a similar substructure in the dark matter
distribution, and attempt to model the mass map with 1-5 mass halos, 
finding a best solution given by 4 mass halos. We note L12's peaks on both maps (blue crosses, denoted as A--D following Ma et al.'s designation).

We also overlay the MEM mass contours on the SL model derived in Section~\ref{sec:sl}, 
shown along with L12 mass peak locations in the bottom-right panel of Figure~\ref{fig:K2D_HST}. 
We note that the WL-MEM results agree very well with our parametric SL results (Section~\ref{sec:sl}), 
both demonstrating an overall agreement between the distributions of the galaxies and DM,
although our parametric SL method does use the light distribution as an initial approximation of the mass.
For the WL-MEM results, 
this is not trivial at all, because the WL-MEM method is entirely non-parametric and is not aware of the distribution of cluster members, 
hence providing a model-independent reconstruction of the underlying total mass distribution.

Although we find an overall agreement with previous and current SL analyses where the DM
follows the location of the galaxies, some  differences are noticeable. 
From the MEM-derived mass map, 
the peak location of the main clump ``C'' is somewhat offset from those of cluster galaxies and X-ray emission
(that rather agree with each other) by about $\sim 0.33\arcmin\approx90$\,kpc$/h$, although their extended features overlap well with each other.
%

The SaWLens derived map, on the other hand, shows much larger differences between the mass and light distributions.
The mass peaks in the SaWLens reconstruction are mostly offset from those of cluster galaxies.
Most notably, the location of the SaWLens mass peak nearest to clump ``A'' is further displaced to the center than the light or the location found by L12, and  clump ``B''  is much less pronounced in this mass map. 

%
The current analysis is limited by the spatial resolution set by the number density of background galaxies,
and may be prone to potential systematics, especially in the critical lensing regime.
In this work we therefore cannot draw any significant conclusion about the level of offsets between the DM, galaxies, and hot gas.

\section{Discussion}
\label{sec:discussion}

\subsection{Mass Profiles from Different Lensing Methods}
\label{subsec:1d2dcomp}

\begin{figure}[tb]
\centering
\includegraphics[width=0.5\textwidth,clip]{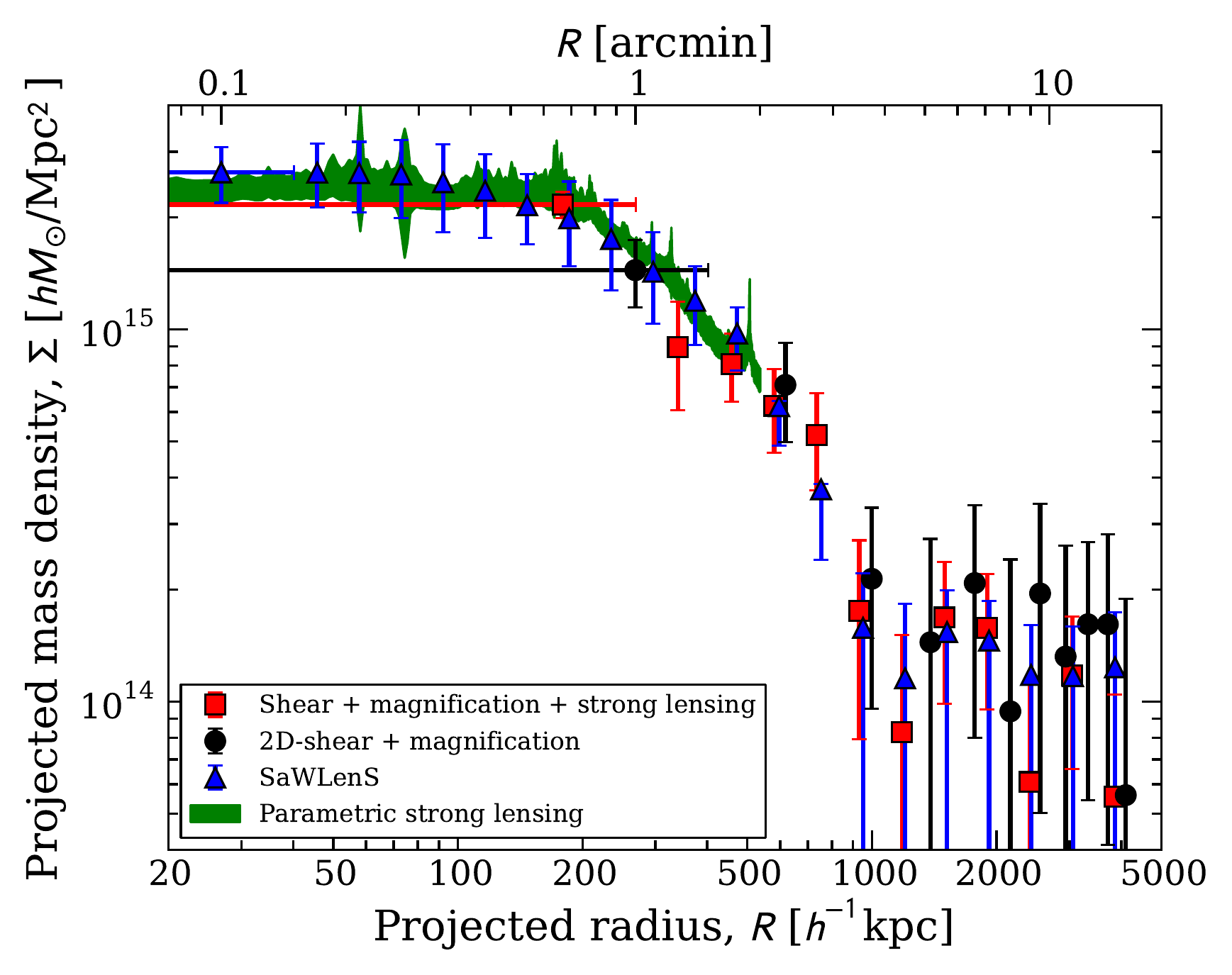}
\caption{Surface mass density profile, combining and comparing results
  from our parametric SL analysis of CLASH-{\it HST} data (green), 
  1D WL (shear+magnification) + SL analysis (red squares), 
  2D WL (shear+magnification) analysis (black circles), and
  SaWLens 2D shear+SL analysis (blue triangles). 
  All the WL-shear analyses are based on both {\it HST} and Subaru datasets,
  where the WL constraints at $\simgt 3\arcmin$ are based on the Subaru observations.
Good
  agreement between the SL and WL analysis is seen in the region of
  overlap. There is also good agreement between the different methods
  of WL analysis.
}
\label{fig:1d-2d}
\end{figure}

The mass profiles we constructed in the previous sections using various methods are summarized in Figure~\ref{fig:1d-2d}: 
a parametric light-traces-mass SL mass profile (green; Section~\ref{sec:sl}),
a non-parametric Bayesian mass profile incorporating WL 1D shear+magnification constraints and the SL mass constraint (red squares; Section~\ref{subsec:1dk}),
a non-parametric approach combining WL 2D shear+ 1D magnification constraints (black circles; Section~\ref{subsec:1dk2dg}),
and finally, a mass profile obtained with SaWLens (blue triangles; Section~\ref{subsec:sawlens}), an independent
non-parametric reconstruction combining SL+WL shear constraints on a
joint multi-scale grid. Here we summarize and compare the different
results.

As can be seen, overall good agreement is found in the regions of overlap
between the different reconstruction methods, within the
errorbars. 
The relatively good agreement we find on large scales is reassuring,
given that it is all based on the same Subaru imaging data. At
$r>1\,\mpch$ there is general change in the sign of the gradient of
the mass profile, where a clear flattening tendency is seen. 
A similar behavior was also found in our recent CLASH lensing analysis of
the high-mass relaxed cluster MACSJ\,1206-08 at $z=0.44$ 
\citep[$M_{\rm vir}\simeq 1.1\times 10^{15}M_\odot/h$,][]{Umetsu+2012}, 
albeit to a lower extent, and may
in both cases reflect the relative prominence of the surrounding
filamentary pattern of structures extending beyond the virial radius,
and hence the relatively less evolved stage of these very massive
clusters at $z\sim0.5$, compared to their lower redshift counterparts,
such as A1689 and A1703, where the outer mass profile continues to
steepen \citep{Broadhurst05b,Umetsu+2011} and for which no
prominent filamentary network is visible.

MACSJ0717 has been recently analyzed by \cite{2012MNRAS.426.3369J}
using multiple {\it HST}/ACS pointings aimed at covering the
filamentary structure around the cluster.
Their mass reconstruction shows slightly higher densities at all radii and
most notably at the outer radii, albeit still in agreement with our 2D
shear+magnification density profile (black circles), within the
errors. This small systematic difference may arise due to Jauzac
et al.'s strategy of surveying only the area where the filamentary
structure is dominant, so that the mean lensing signal that we obtain with our
larger FOV is on average expected to be lower, as observed.

The total masses of the cluster
are presented in Section~\ref{subsec:lcdm} where we compare
with prediction for the highest
mass halos expected in the context of the \lcdm\ model.

\subsection{X-Ray results from Chandra and XMM-Newton}\label{subsec:xray}

Here we compare complementary X-ray derived total masses to our 1D
lensing-derived total masses.
One mass estimator uses the assumption that the gas 
fraction at $r_{2500}$ is constant (for this work we assume $f_{\rm gas}\equiv M_{gas}/M_{total}=0.11$).
The other common mass estimator uses the assumption that
the primary pressure support for the gas is thermal pressure, that
is, that the gas is in nearly hydrostatic equilibrium (HSE). While
it is true that this system is merging and highly complex \citep[as shown
in][]{Ma2009}, a comparison of the HSE mass with the lensing mass
can be used to place estimates on the magnitude of non-thermal pressure
support in the gas, such as bulk motions and turbulence.

We analyze public Chandra observations from a single OBS-ID (4200) with a net 
exposure time of 59,000\,seconds. Gas density profiles from the ACCEPT database \citep{Cavagnolo08} were used
to generate an enclosed gas mass profile. We also derived radial profiles based on spectra extracted from concentric annular apertures.
The tool used for the HSE mass estimate is JACO \citep[Joint Analysis of Cluster 
Observations,][]{JACO}; we refer the reader to this paper for the
details of the X-ray analysis procedure. 
The best-fit NFW parameters derived from the joint JACO fit, at $r_{2500}$ are,
$M_{2500} = 4.45^{+0.95}_{-0.46} \times 10^{14}\,M_\sun/h$, 
$c_{2500} =  0.6\pm0.3$.
The X-ray profile has a limited range $r<0.7\,\mpch$. 

We compare the X-ray total mass within $r<500\,\kpch$ with lensing-derived values at the same clustercentric radius in Table~\ref{tab:nfw}. The X-ray mass seems to be about $\sim87\%$ of the lensing 1D WL+SL NFW fitted value.
In Figure~\ref{fig:SZ} we compare the resulting total mass ($M_{\rm 3D}<r$) profiles from the X-ray fit (magenta) and the lensing 1D WL+RE NFW fit (orange).
We find that the X-ray mass profile derived with the HSE assumption underestimates the lensing profile at all radii. The difference is most extreme in
the core of the cluster, where the HSE X-ray mass is only $45\%$ of
the WL mass, rising to agreement with the WL mass
at the limit of the X-ray data, $r=0.7\mpch$.
This inconsistency clearly stems from the departure from
HSE at the center of the cluster where significant merging
activity has been claimed, augmented by the highly
non-spherical shape of the cluster \citep{Meneghetti2010,Rasia2012}.
We enhance our discussion below by further comparing to the SZE-derived mass (see Section~\ref{subsec:sze}) which also relies on the HSE assumption.

For a wider look at the 2D distribution of the gas, we also analyze XMM-Newton observations of MACSJ0717 recently made public
(P.I. Million, observations 672420101, 672420201, 672420301, totaling
145,000\,seconds). We perform a standard reduction on the observations,
removing high flare times. To achieve the best sensitivity to gas at
large clustercentric radius, we summed the MOS1, MOS2, and p-n CCDs of all three
observation sets. The brightest point sources were identified and
excluded.  Since the PSF for XMM gets fairly
distorted and elongated at off-axis angles, there is incomplete
removal of the halos around point sources there.

\begin{figure}[tb]
\centering
\includegraphics[width=0.55\textwidth,angle=0,clip]{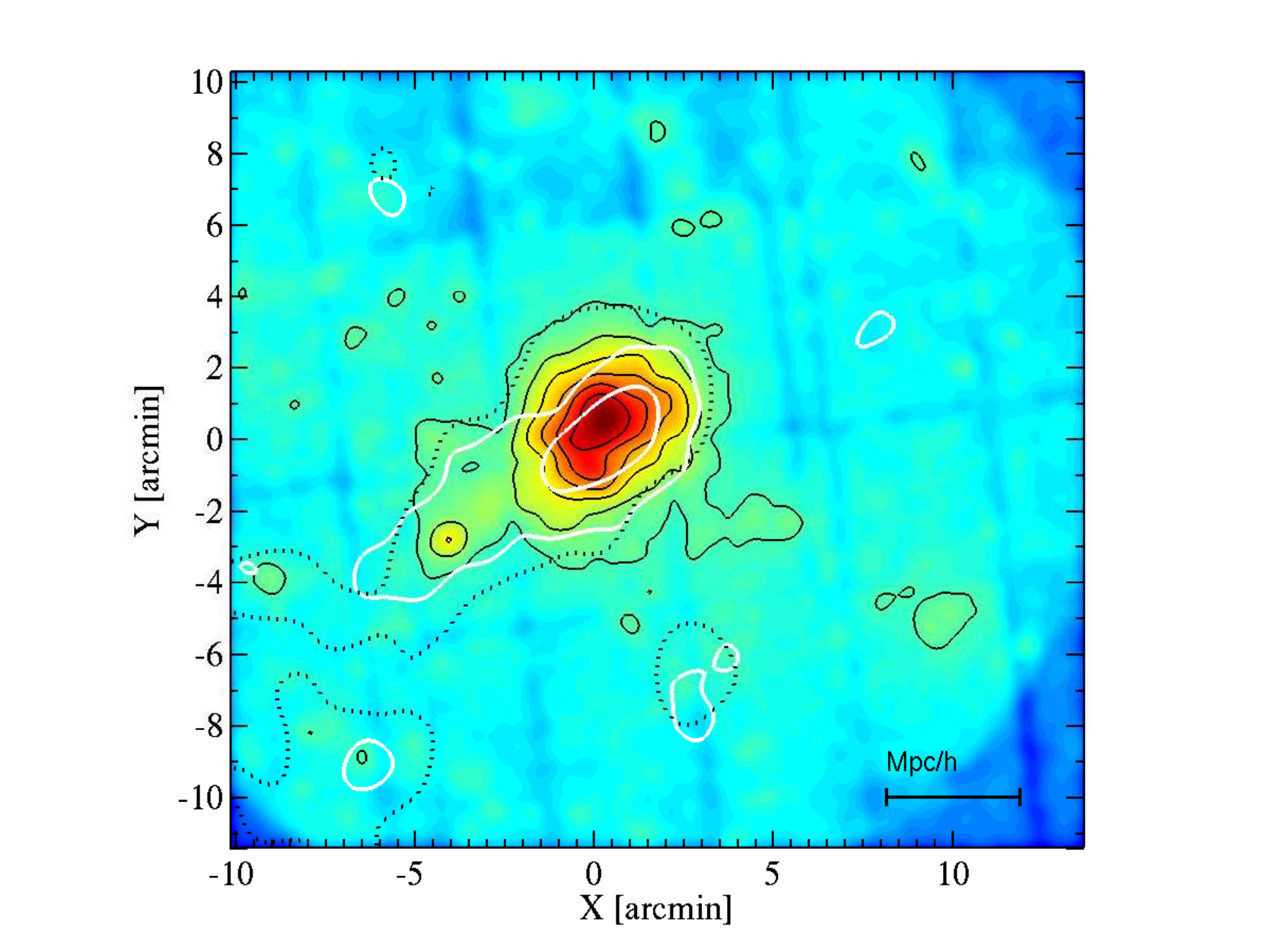}
\caption{X-ray map from XMM-Newton data in logarithmic scale, smoothed to $\theta_{\rm FWHM}=0.5\arcmin$, with $1.5\sigma$ contours with $2\sigma$ intervals (solid black lines). Also overlaid are $\kappa$ contours (solid white lines) and galaxy density contour (dotted black line). }
\label{fig:X}
\end{figure}

We present the logarithmic X-ray brightness map in Figure~\ref{fig:X}, 
where we smooth the map with a Gaussian of $\theta_{\rm FWHM}=0.5\arcmin$ 
and overlay it with logarithmic contours starting at $1.5\sigma_X$ above the 
background with $2\sigma_X$ intervals (solid black lines). For comparison, 
we overlay the mass density (2.5- and 6.5-$\sigma_\kappa$) contours from the Subaru lensing map (solid white line) and the galaxy distribution 2.5$\sigma_n$ contour (dotted black line). 

The cluster X-ray brightness peak is clearly shifted from the mass (and galaxy) 
peak, at about $R=0.37\arcmin= 100\,\kpch$ north-west of the optical center.
This has been noted in previous detailed analyses of the cluster center \citep{Ma2009} and was interpreted to be
due to trailing of the gas component behind the cluster as subclumps
cross through the main cluster halo, because the gas and
the DM have different cross-sections.
We summarize in Table~\ref{tab:cluster} the location of the optical center we have 
been using and that of the X-ray brightness peak. 

With the advent of the XMM-Newton data, we can also see here the clear detection of the secondary massive component (denoted $z_5$ in our mass map, see Figure~\ref{fig:kappa2D}) at a $5.5\sigma_X$ level, and the third mass component (denoted $z_2$ in our mass map) at a $2.5\sigma_X$ level. Finally, the forth mass component detected from lensing, $z_3$, is evident in the map but at much lower significance,  $\sim1.6\sigma_X$. Although lensing shows this halo to be more massive than $z_2$, it is hard to detect it in the X-ray image since it lies at the edge of our data where it is more prone to contamination by point sources.

Overall, it seems the gas in the outskirts (at a scale of $\simlt0.7\mpch$)
is more spherically-distributed around the main cluster component,
but has the same substructure corresponding
to high-mass peaks along the filamentary structure.  The emission from the hot 
gas is expected to be more spherical in shape than that mapped by lensing because
the gas in projection traces the gravitational potential, which is rounder than 
the density distribution.

\subsection{SZE Results using  Bolocam}\label{subsec:sze}

We obtained SZE observations of MACSJ0717 using Bolocam at the
Caltech Submillimeter Observatory.  Bolocam is a 144-element
bolometric array operating at 140~GHz with an angular resolution of
$58\arcsec$ (full-width at half-maximum).  The data were collected by
scanning Bolocam's $8\arcmin$ FOV in a Lissajous pattern to
produce an image with coverage extending to $\simeq 10\arcmin$ in
radius.  Full details of the observations are given in
\cite{2012ApJ...761...47M} and \cite{sayers2012}, and the data were
reduced using the procedures described in \cite{2011ApJ...728...39S}.
In addition, we note that in the multiwavelength SZE analysis of
\cite{2012ApJ...761...47M}, which used the same Bolocam SZE data as
this analysis, a tentative kinetic SZE signal from subclump ``B'' of
\cite{Ma2009} is reported 
at a significance of $\simeq 2\sigma$.  Consequently, we
subtracted the best-fit kinetic SZE template of
\cite{2012ApJ...761...47M} from our data, although we note that
\cite{2012ApJ...761...47M} were not able to robustly constrain a
spatial template of this kinetic SZE signal, modeling it as a simple
Gaussian centered on the location of ``B''.  We then
fit a spherical generalized NFW pressure (gNFW) model \citep{Nagai07}
to our SZE data allowing both the normalization and scale radius to
vary while using the best-fit slope parameters found by
\cite{arnaud2010}.

\begin{figure}[tb]
\centering
\includegraphics[width=0.5\textwidth,angle=0,clip]{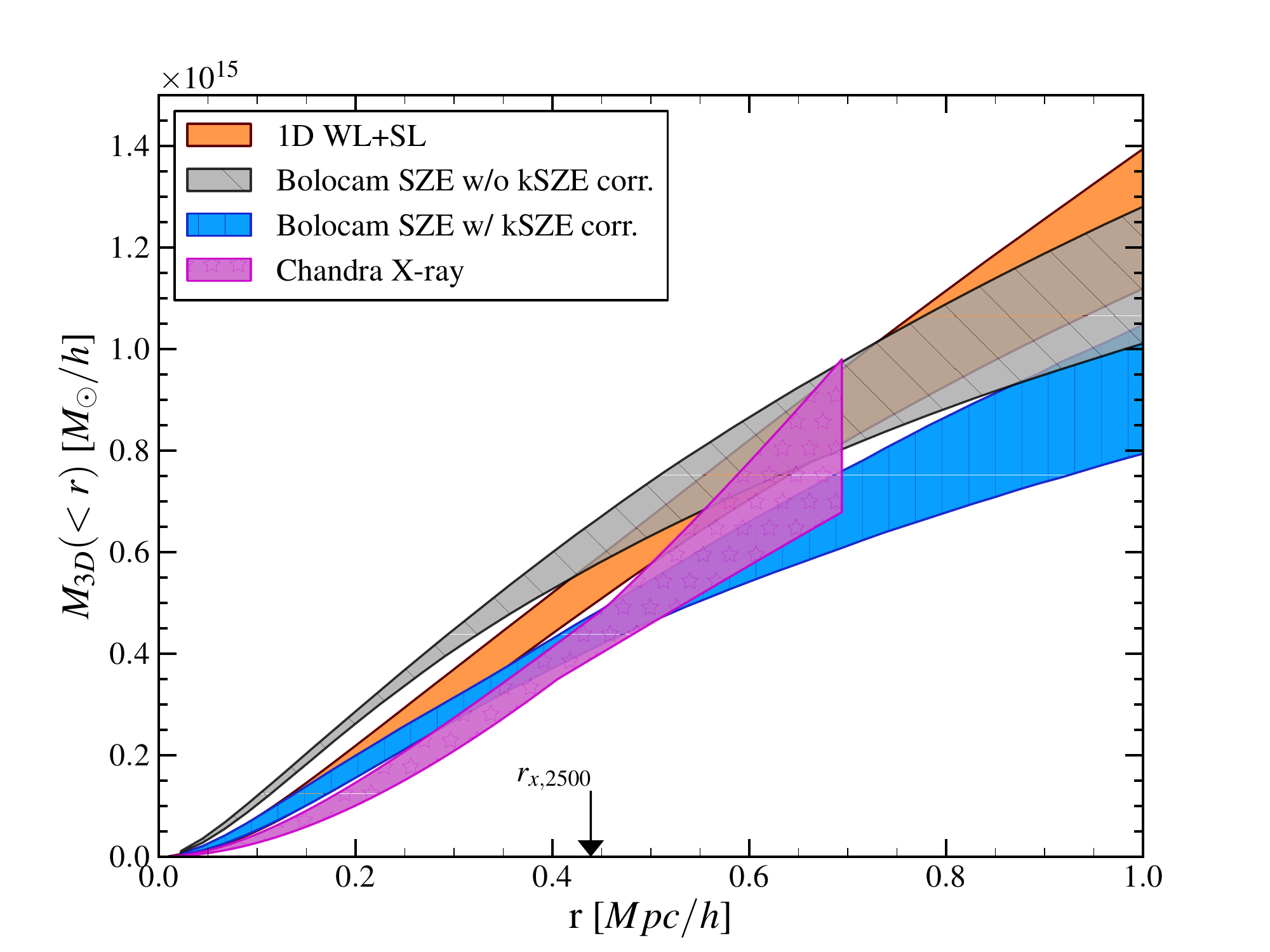}
\caption{Integrated total mass profiles $M_{\rm 3D}(<r)$, derived from an NFW
  fit to WL+SL mass reconstruction (orange), from the
  SZE data after subtracting the best-fit kinetic SZE template found
  by \cite{2012ApJ...761...47M} (blue) and from X-ray (magenta). The
  SZE and X-ray fits are largely consistent with each other, but
  systematically lower than the lensing derived mass profile at all
  radii, presumably due to biases associated with the HSE assumption used
  in deriving both the SZE and X-ray mass profiles.  For completeness,
  we also show the mass profile derived from the SZE data without
  correcting for the kinetic SZE signal (gray).}
\label{fig:SZ}
\end{figure}

We compute the mass of MACSJ0717 from this gNFW fit to the SZE data
using the formalism described by \cite{Tony2011}.  In particular, we
assume that the cluster is spherical and in HSE, although we note that
neither of these assumptions are strictly valid for MACSJ0717 due to
its complex dynamical state.    The
resulting SZE mass profile is shown in blue in Figure~\ref{fig:SZ},
and we note that it is approximately $\sim75$\% of our
lensing-derived mass profile.  However, the SZE mass profile is in
reasonably good agreement with the X-ray derived mass profile, which
also assumed HSE.  We therefore conclude that the difference is                       
due to the bias associated with assuming HSE. For example, even for 
relaxed systems we
expect the HSE-derived virial mass to be biased approximately
$10-20$\% low on-average compared to the true virial mass due mainly
to bulk motions in the gas \citep{Nagai07,Rasia2004,Rasia2012,Meneghetti2010}. Furthermore, in order
to gauge the impact of the kinetic SZE signal from subclump ``B'', we
also computed an SZE mass profile without subtracting the kinetic SZE
template (shown in gray in Figure~\ref{fig:SZ}).  As expected, the
difference between the two SZE mass profiles is most significant in
the inner regions of MACSJ0717 near the center of subclump ``B'', where the corrected mass is $\sim50\%$ lower than the uncorrected one,
decreasing to only $\lesssim 15$\% difference at large radius.

\subsection{Total Mass Compared with \lcdm\ Predictions}
\label{subsec:lcdm}


\begin{figure}[t]
 \hspace{-1.5cm}
\includegraphics[width=0.65\textwidth,clip]{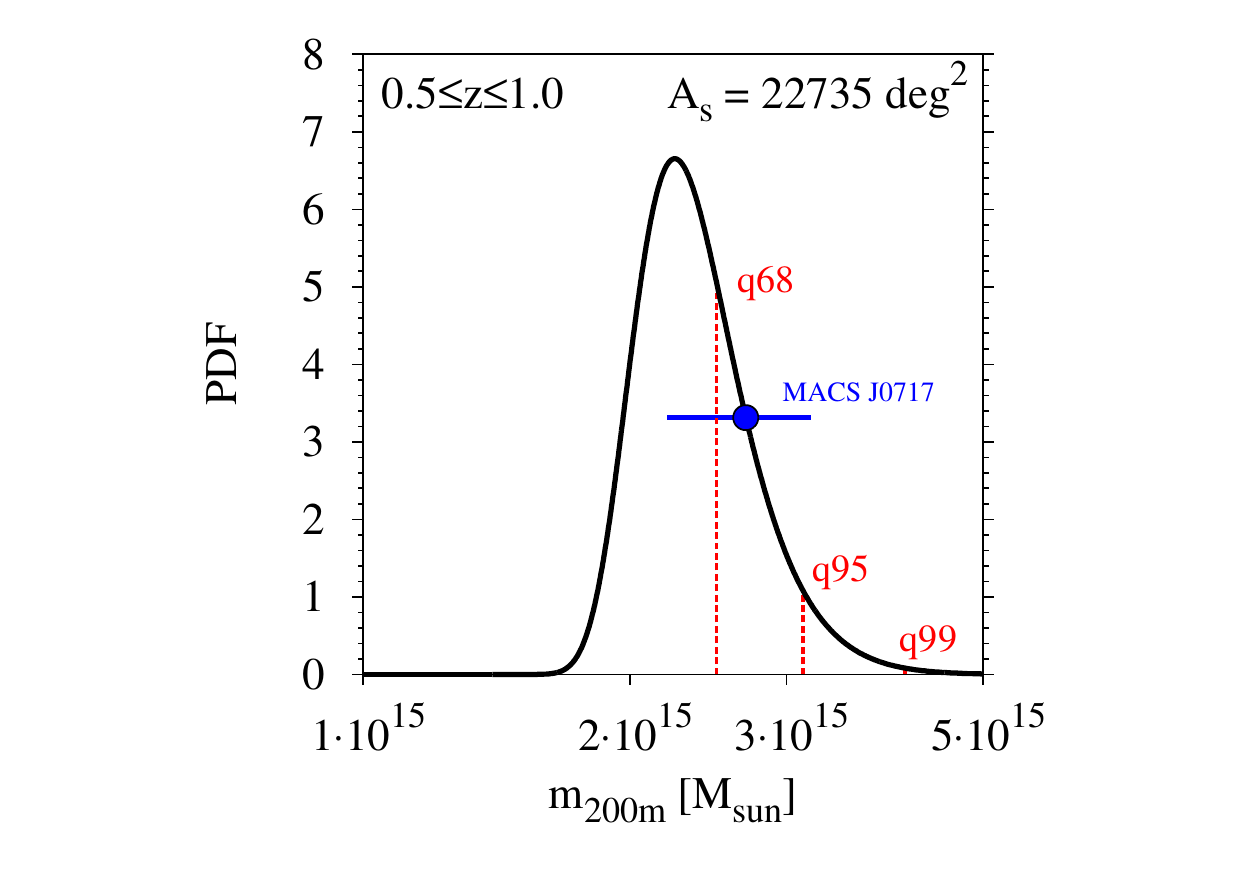}
\caption{PDF of the maximum cluster mass expected to be found in a survey with sky coverage of the MACS survey, $A_S=22,735\deg^2$, within the redshift interval $0.5\leq z\leq1.0$. The red vertical lines represent the $1,2,3\sigma$ upper confidence levels. The blue circle denotes the measured lensing mass of MACSJ0717.}
\label{fig:PDF}

\end{figure} 

We have reliably estimated the total mass of the cluster in Sections~\ref{subsec:1dk} and~\ref{subsec:1dk2dg} to be $\mvir({\rm 1DWL+SL})=2.13\times10^{15}\,M_sun/h$ and $\mvir({\rm 2DWL})=2.23\times10^{15}\,M_sun/h$, respectively, as summarized in Table~\ref{tab:nfw}. We note that the two masses are in excellent agreement. In Section~\ref{subsec:2Dhalo}, we derived specific masses for the individual halos comprising the cluster and its surrounding stuctures, in order to more accurately account for the LSS contribution to the mass.
Here we use as the total mass for MACSJ0717 the
sum of the virial masses of the two components that lie
within the virial region of the main halo, $\mvir(z_5+z_6)\approx(1.96\pm0.28)\times\mhunit$, as determined by multi-halo
modeling (see Tables~\ref{tab:nfwMH}) which is slightly lower, but still consistent
with the values found from single-halo modeling noted above.

Another total mass estimate for this cluster was previously presented in \cite{2012MNRAS.427.1298H}, part of a larger 50-cluster batch analysis. That measurement solely relied on WL shear measurement, and did not simultaneously constrained both $\cvir$ and $\mvir$, as done here, but relied on the \cite{Duffy+2008} relation, which according to their published mass of $\mvir=5 \times10^{15} M_\sun/h_{70}$ should be $\cvir=3.05$.  Forcing this concentration value, our mass would have been $\mvir=(4.5\pm0.6)\times10^{15} M_\sun/h_{70}$, consistent with their mass. Since there is a strong negative correlation between the mass and concentration parameters, such an assumption of a low concentration will lead to an overestimated high mass.


We are now in the position to address 
the probability of finding such a massive cluster in our Universe via extreme 
value statistics \citep{2011PhRvD..83b3015M,2011MNRAS.414.2436C,2012MNRAS.420.1754W,Hotchkiss11,Harrison2012}.
In the work of \cite{2012MNRAS.420.1754W} several high-mass clusters
were examined (Abell\,2163, Abell\,370, RXJ\,1347$-$1145 and
1E0657$-$558) and the cumulative probability function of finding
such massive systems in their given survey area was calculated using
general extreme value statistics. However, they found that none
of those clusters alone is in tension with \lcdm. In another paper,
\cite{2012A&A...547A..67W} test the probability of the extremely large
Einstein radius measured for MACSJ0717, previously estimated as $\theta_E=55\arcsec$ \citep[for $z=2.5$,][]{Zitrin+2009_0717}
within
the standard \lcdm\ cosmology. Although they find this system not to be
in tension with the \lcdm\ expectations, the sensitivity of the Einstein radius distribution to
various effects (lens triaxiality, mass-concentration, inner slope of
the halo density profile and more) implies that this comparison is much less
reliable than that using a total lensing mass. To conclude,
\cite{2012A&A...547A..67W} calculate that a galaxy cluster in the
redshift range $0.5\leq z\leq1.0$ would need to have a total mass of at least
$M_{200m} = 4.5\times10^{15}\,M_\odot$, where $M_{200m}$ is defined with respect to the mean density of the universe at the time of collapse, in order to exclude \lcdm\ at the
$3\sigma$ level.

To
compare with Waizmann et al.'s values,  MACSJ0717 has $M_{200m}\approx
(2.9\pm0.5)\times10^{15}\,M_\odot$ from lensing. In Figure~\ref{fig:PDF} we show the probability distribution function (PDF) of the most massive cluster \citep[for details of this calculation see][]{2011MNRAS.413.2087D,2011MNRAS.418..456W,2012MNRAS.420.1754W}
as calculated for the MACS survey area, $A_S=22,735\deg^2$, within the a-priori 
redshift interval $0.5\leq z\leq1.0$. We overlay the total mass estimate of 
MACSJ0717 as the blue error bar. It can be seen that the mass of MACSJ0717 falls 
slightly below the $95$ per cent quantile. In order to avoid the bias in the occurrence 
probabilities of rare galaxy clusters as discussed in \cite{Hotchkiss11}, one has to 
set the redshift interval a-priori. Since MACSJ0717 happens to fall at the lower 
end of the chosen redshift interval, the PDF in Figure~\ref{fig:PDF} can be considered as 
a conservative estimate of the true rareness of the cluster such that a more thorough 
treatment, as presented in \cite{2012arXiv1210.4369H}, would yield a lower 
rareness. 

We therefore conclude that MACSJ0717's mass, being the largest above $z=0.5$, does not 
pose any tension on the standard \lcdm\ cosmology. However, the growing number 
of very massive clusters at high redshifts makes it advisable to study the 
occurrence probabilities of ensembles of massive clusters instead of single ones 
 \citep[see e.g.,][]{2012arXiv1210.6021W}.

\section{Summary and Conclusions}
\label{sec:summary}

In this paper, we have presented a comprehensive lensing analysis
of the merging cluster MACSJ0717 at $z=0.5458$,
the largest known cosmic lens with complex internal structures,
by combining independent constraints from WL distortion,
magnification, and SL effects.
This is based on wide-field Subaru+CFHT $u^*BVR_{\rm c}i'z'JK_{\rm S}$ imaging, 
combined with detailed deep CLASH-{\it HST} 16-band imaging in the cluster core.

We have obtained an improved SL re-analysis of the inner mass distribution from the CLASH {\it HST} data and confirmed our earlier model \citep{Zitrin+2009_0717}, where we now measure the effective Einstein radius of $60\pm6\arcsec$ ($z=2.963$) in agreement with previous values.

The deep Subaru multi-band photometry is used to separate background,
foreground, and cluster galaxy populations 
using the multi-color selection techniques established in our earlier work
\citep{Medezinski+2010,Umetsu+2010_CL0024}, allowing us to obtain a
reliable WL signal free from significant contamination of
unlensed cluster and foreground galaxies for this relatively high
redshift cluster.  By combining complementary SL, WL distortion and
magnification measurements, we have constructed a model-free mass
distribution out to well beyond the virial radius ($r_{\rm vir}\approx
2\,$Mpc\,$h^{-1}$), effectively breaking the mass-sheet degeneracy.

 By fitting projected NFW profiles to the reconstructed mass profiles,
we obtain consistent virial mass estimates:
$\mvir = (2.13\pm0.47)\times 10^{15}\,M_{\odot}/h$ from the joint likelihood analysis of tangential shear, magnification, and SL measurements
and 
$\mvir = (2.23\pm0.41)\times 10^{15}\,M_{\odot}/h$ from the joint likelihood analysis of 2D shear and azimuthally-averaged magnification measurements.

We further preformed a 2D lensing analysis to compare the
reconstructed total mass distribution with the luminosity and galaxy
distributions over a wide Subaru FOV.  The Subaru data reveal
the presence of several mass components that lie along a filamentary
structure at the cluster redshift. 
We modeled the 2D shear field with a composite of 
nine NFW halos in projection,
corresponding to mass peaks detected above $2.5\sigma$,
including the main cluster approximated by an elliptical NFW halo.
Those mass halos, which are part of the $z\approx0.55$ structure, seem to form a
filamentary-like structure that spans $4$\,Mpc/$h$.  The sum of the two halos that lie within the virial radius, $\mvir(z_5+z_6)\approx(1.96\pm0.28)\times\mhunit$, is in good agreement with the total mass estimates from the single-NFW fits to the two joint likelihood analyses we have preformed.
We conclude that the four mass halos,
which are confined to the volume within $\approx 2\rvir$, i.e. the turn-over radius,   
will likely accrete onto the main cluster to form an even more massive system by $z=0$, 
as much as $\mvir\approx2.5\times\,\mhunit$ ($\simeq 3.6\times 10^{15}\,M_\odot$). 

Utilizing the high-resolution {\it HST}-CLASH observations, we were able to resolve the inner core region of the cluster 
using two independent non-parametric methods, one relying on the WL and one combining both WL+SL constraints. 
Both show a good overall agreement between the light and the total mass distribution.
However, the current spatial resolution is not enough to precisely measure the offsets (if any) between DM, galaxy, and X-ray centroids.
The upcoming ``Frontier Fields'' initiative with HST will target 6 massive lensing clusters, 
MACSJ0717 amongst them, and will yield unparalleled deep and detailed mass maps of this complicated merging cluster, 
which will allow for the first time to significantly determine the possible offsets between the mass components of MACSJ0717, and improve current constraints on the collisional nature of DM.

Using extreme value statistics,
we found our mean mass estimate of $\mvir\approx2\times 10^{15}\,
M_{\odot}/h$ does not lie outside \lcdm\, predictions, for which a higher mass of at least $\mvir\approx3.5\times
10^{15}\, M_{\odot}/h$ is required for \lcdm\ to be ruled out at the $3\sigma$
level, but instead differs only at the 
$2\sigma$ level, so that larger samples of clusters are motivated 
by our findings, to provide a more definitive joint probability.

In conclusion, we find MACSJ0717 to be the most massive cluster so far
detected at $z>0.5$, and to exhibit significant substructure not only
in the center, but also to be part of a LSS that spans
over $\sim\,4\mpch$, accreting nearby clusters themselves as massive as
few$\times10^{14}\,M_\sun/h$. This prominence of surrounding structure 
that we find in our work, including subclusters and filaments 
together with the ongoing merging activity found in the core of
MACSJ0717 indicate that this object is relatively unevolved compared to
massive clusters at lower redshift where relaxed clusters such as A1689 show
little surrounding structures.


\acknowledgments

We acknowledge useful discussions with Nobuhiro Okabe, Assaf Horesh, Omer Bromberg and Preethi Nair.
We thank Nick Kaiser for making the IMCAT package publicly available.  
We thank G. Mark Voit for having contributed to the ACCEPT-based
X-ray mass measurements in advance of publication. We are greateful to Zoltan Levay who crafted the HST color image.
We are greateful to Jason Rhodes and Richard Massey for providing us with the RRG package. 

The CLASH Multi-Cycle Treasury Program is based on observations made
with the NASA/ESA {\em Hubble Space Telescope}. 
The Space Telescope Science Institute is operated by
the Association of Universities for Research in Astronomy, Inc.~under
NASA contract NAS 5-26555. 
ACS was developed under NASA contract NAS 5-32864.
EM is supported by NASA grant {\it HST}-GO-12065.01-A.
KU acknowledges partial support from the National Science Council of Taiwan (grant NSC100-2112-M-001-008-MY3) and from the Academia Sinica Career Development Award.
The Bolocam observations were partially supported by the Gordon and Betty
Moore Foundation. 
This material is based upon work at the Caltech Submillimeter Observatory, which is operated by the California Institute of Technology under cooperative agreement with the National Science Foundation (AST-0838261).
JS is supported by NSF/AST0838261 and NASA/NNX11AB07G;
NC is partially supported by a NASA Graduate Student Research Fellowship.
AZ is supported by the ``Internationale Spitzenforschung II/2'' of the
Baden-W\"urttemberg Stiftung.
TM is provided by NASA through the Einstein Fellowship
Program, grant PF0-110077.
This research was carried out in
part at the Jet Propulsion Laboratory, California Institute
of Technology, under a contract with NASA.
MN acknowledges support from PRIN-INAF 2010.


\bibliography{A1689,papers,KeiichiU,Lensref,CMB,Elinor}



\begin{deluxetable}{lc}
\tablecolumns{2}
\tablecaption{
 \label{tab:cluster}
Properties of the galaxy cluster MACSJ0717
} 
\tablewidth{0pt}  
\tablehead{ 
 \multicolumn{1}{c}{Parameter} &
 \multicolumn{1}{c}{Value} 
} 
\startdata
ID .............................................. &  MACS J0717.5+3745 \\
Optical position (J2000.0) & \\
\ \ \ \ R.A. ...................................... & 07:17:32.63\\
\ \ \ \ Decl. ..................................... & +37:44:59.7\\
X-ray peak position (J2000.0) & \\
\ \ \ \ R.A. ...................................... & 07:17:31.65    \\
\ \ \ \ Decl. ..................................... & +37:45:18.5 \\
Redshift .................................... & $0.5458$\\
X-ray temperature  (keV) ..........& $12.5\pm 0.7$\\
Einstein radius ($\arcsec$) ....................& $60\pm 3$
at $z_s=2.963$
\enddata
\tablecomments{
The cluster MACS\,J0717.5+3745($z = 0.5458$) was discovered in the
Massive Cluster Survey (MACS) as described by Reference [1], and its redshift determined by [2].
The optical cluster center is defined as the center of the bright red-sequence selected galaxies. The X-ray center and mean temperature were taken from Reference [3]. We note, the mean temperature can vary by larger than the quoted uncertainty between authors since it depends on the exact location of the X-ray center, which is different between [1] and [3].
}
\tablerefs{ 
 [1] \cite{Ebeling+2001_MACS};
 [2] \cite{Ebeling+2007};
 [3] \cite{Postman+2012_CLASH};
 }
\end{deluxetable}


\begin{deluxetable}{cccc}
\tablecolumns{4}
\tablecaption{
 \label{tab:subaru}
Subaru/Suprime-Cam + CFHT/MegaPrime data.
} 
\tablewidth{0pt} 
\tablehead{ 
 \multicolumn{1}{c}{Filter} &
 \multicolumn{1}{c}{Exposure time\tablenotemark{a}} &
 \multicolumn{1}{c}{Seeing\tablenotemark{b}} &
 \multicolumn{1}{c}{$m_{\rm lim}$\tablenotemark{c}}  
\\
 \colhead{} &
 \multicolumn{1}{c}{(ks)} &
 \multicolumn{1}{c}{(arcsec)} &
 \multicolumn{1}{c}{(AB mag)} 
} 
\startdata  
 $u^*$ & 19.63 & 0.94 & 26.1\\
 $B$  & 3.84 & 0.95 & 26.6\\
 $V$  & 2.16 & 0.69 & 26.4\\
 $\RC$\tablenotemark{d}  & 2.22 & 0.79 & 26.1 \\ 
 $i'$  & 0.45 & 0.96 & 25.4 \\
 $z'$  & 5.87 & 0.85 & 25.6 \\
 $J$  & 0.9 & 0.73 & 22.7 \\
 $K_{\rm S}$  & 0.5 & 0.54 & 23.0 
\enddata
\tablenotetext{a}{Total exposure time.} 
\tablenotetext{b}{Seeing FWHM in the full stack of images.}
\tablenotetext{c}{Limiting magnitude for a $3\sigma$ detection within a
 $2\arcsec$ aperture.}
\tablenotetext{d}{Band used for WL shape measurements.}
\end{deluxetable}


\begin{deluxetable}{ccccc}
\tabletypesize{\footnotesize}
\tablecolumns{5} 
\tablecaption{
 \label{tab:color}
Galaxy color selection.
}  
\tablewidth{0pt}  
\tablehead{ 
 \multicolumn{1}{c}{Sample} &
 \multicolumn{1}{c}{Magnitude limits\tablenotemark{a}} &
 \multicolumn{1}{c}{$N$} &
 \multicolumn{1}{c}{$n_g$\tablenotemark{b}} & 
 \multicolumn{1}{c}{$\langle z_s\rangle$\tablenotemark{c}} 
\\
 \colhead{} & 
 \multicolumn{1}{c}{(AB mag)} &
 \colhead{} &
 \multicolumn{1}{c}{(arcmin$^{-2}$)} &
 \colhead{}
} 
\startdata  
 Red      & $21<z'<25$ &  10490 & 9.6 & 1.21 \\
 Green    & $z'<24$      &  1252  & 1.3 & 0.49/0.54 \\
 Blue     & $22.5<z'<26$ &  11998  & 11.5 & 2.23 
\enddata 
\tablenotetext{a}{Magnitude limits for the galaxy sample.}
\tablenotetext{b}{Mean surface number density of source background galaxies.} 
\tablenotetext{c}{Mean photometric redshift of the sample obtained with the BPZ
code.}
\end{deluxetable}


\begin{deluxetable}{cccccccc}
\tabletypesize{\footnotesize}
\tablecolumns{8} 
\tablecaption{
 \label{tab:wlsamples}
Galaxy samples for WL shape measurements.
}  
\tablewidth{0pt} 
\tablehead{ 
 \multicolumn{1}{c}{Sample} &
 \multicolumn{1}{c}{$N$} &
 \multicolumn{1}{c}{$n_g$\tablenotemark{a}} & 
 \multicolumn{1}{c}{$\sigma_g$\tablenotemark{b}} & 
 \multicolumn{2}{c}{$z_{s,{\rm eff}}$\tablenotemark{c}} &
 \multicolumn{2}{c}{$\langle D_{ls}/D_s\rangle$\tablenotemark{d}} 
\\
 \colhead{} & 
 \colhead{} &
 \multicolumn{1}{c}{(arcmin$^{-2}$)} &
 \colhead{} &
 \multicolumn{1}{c}{\scriptsize MACSJ0717} &
 \multicolumn{1}{c}{\scriptsize COSMOS} &
 \multicolumn{1}{c}{\scriptsize MACSJ0717} &
 \multicolumn{1}{c}{\scriptsize COSMOS} 
} 
\startdata  
 Red      & 4856 & 5.65 &  0.41 & 
1.1 & 1.08 & 
0.43 & 0.42\\
 Blue    &  4738 & 5.6 & 0.42 & 
1.89  & 1.56 & 
0.59 & 0.55\\
 Blue+red & 9594 & 11.2 & 0.41 & 
1.26 & 1.27 & 
0.48 & 0.48
\enddata 
\tablecomments{
}
\tablenotetext{a}{Mean surface number density of galaxies.} 
\tablenotetext{b}{Mean RMS error for the shear estimate per galaxy, 
 $\sigma_g\equiv (\overline{\sigma_g^2})^{1/2}$.}
\tablenotetext{c}{Effective source redshift corresponding to the mean
 depth $\langle\beta\rangle$ of the sample.}
\tablenotetext{d}{Distance ratio 
 averaged over the redshift distribution of the sample,
 $\langle\beta\rangle$.}
\end{deluxetable}

 
\begin{deluxetable}{lcccccc}
\tabletypesize{\footnotesize}
\tablecolumns{7}
\tablecaption{\label{tab:nfw} Best-fit NFW parameters to non-parametric mass reconstruction\\ and comparison with X-ray and SZE masses}
\tablewidth{0pt}
\tablehead{
\multicolumn{1}{l}{Method                    }&
\multicolumn{1}{c}{$R_{\rm vir}$             }&
\multicolumn{1}{c}{$M_{\rm vir}$             }&
\multicolumn{1}{c}{$M(<500\,\kpch)$                }&
\multicolumn{1}{c}{$\kappa_{\rm c}$ \tablenotemark{a}}&
\multicolumn{1}{c}{$M_{\rm vir}/L_{\RC}$}&
\multicolumn{1}{c}{$\chi^2/$DOF\tablenotemark{b}} 
\\ 
\multicolumn{1}{l}{                          }&
\multicolumn{1}{c}{(Mpc$/h$)           }&
\multicolumn{1}{c}{($10^{15}M_\odot/h$)}&
\multicolumn{1}{c}{($10^{15}M_\odot/h$)}&
\multicolumn{1}{c}{                          }&
\multicolumn{1}{c}{($h\,M_\odot/L_{\odot}$)  }&
\multicolumn{1}{c}{                          }
}
\startdata 

1D WL+SL\tablenotemark{c} &$1.94\pm0.14$ & $2.13^{+0.49}_{-0.44}$ & $0.63\pm0.17$ & $<0.01$ &$301\pm66$ & 10/9\\
\hline \\[-5pt]

2D WL\tablenotemark{d} &$1.97\pm0.15$ & $2.23^{+0.44}_{-0.38}$ & $0.54\pm{0.12}$ & $0.03\pm0.01$ &$310\pm57$ & 25/7\\

\hline \\[-5pt]
X-ray & & & $0.54 ^{+0.04} _{-0.08}$&  & & \\
\hline \\[-5pt]
SZE & & & $0.50 \pm 0.04$&  & & 
\enddata
\tablecomments{
\tablenotetext{a}{The constant dimensionless mass-sheet component.}
\tablenotetext{b}{The goodness-of-fit, minimized $\chi^2$ over number of DOF.}
\tablenotetext{c}{The NFW model was fitted to the mass profile fully reconstructed from WL+SL (see \S~\ref{subsec:1dk}), constrained by 1D WL (shear+magnification)+SL Einstein-radius. }
\tablenotetext{d}{here, The NFW model was fitted to the mass profile reconstructed from WL alone (see \S~\ref{subsec:1dk2dg}), using  WL (2D shear+1D magnification), without SL constraints.}
}
\end{deluxetable}

 
\begin{deluxetable}{lccccccccc} 
\tabletypesize{\footnotesize}
\tablecolumns{10}
\tablecaption{\label{tab:nfwMH} Best-fit NFW Multi-halo parameters of the 2D shear Analysis} 
\tablewidth{0pt} 
\tablehead{ 
\multicolumn{1}{l}{Halo} &
\multicolumn{1}{c}{$\Delta X$\tablenotemark{a}} &
\multicolumn{1}{c}{$\Delta Y$\tablenotemark{a}} &
\multicolumn{1}{c}{$z_{\rm l}$\tablenotemark{b}} &
\multicolumn{1}{c}{$z_{\rm s}$\tablenotemark{c}} &
\multicolumn{1}{c}{$R_{\rm vir}$} &
\multicolumn{1}{c}{$M_{\rm vir}$\tablenotemark{d}} &
\multicolumn{1}{c}{$e$\tablenotemark{e}} &
\multicolumn{1}{c}{$\theta_e$\tablenotemark{f}} &
\multicolumn{1}{c}{$M_{\rm vir}/L_{\RC}$} 
\\ 
\multicolumn{1}{l}{} &
\multicolumn{1}{c}{(arcmin)} &
\multicolumn{1}{c}{(arcmin)} &
\multicolumn{1}{c}{}  &
\multicolumn{1}{c}{} &
\multicolumn{1}{c}{(Mpc$/h$)} &
\multicolumn{1}{c}{($10^{15}M_\odot/h$)} &
\multicolumn{1}{c}{}  &
\multicolumn{1}{c}{(deg)} &
\multicolumn{1}{c}{($h\,M_\odot/L_{\odot}$)} 
}
\startdata 

z1 & -12.11 & 6.23 & 0.29 & 1.21 & 1.45& $0.58\pm0.16$   &  $-$ & $-$ & $230\pm63$ \\ 
z2 & -9.73 & -3.57 & 0.56 & 1.27 & 0.79& $0.15\pm0.09$   &  $-$ & $-$ & $243\pm142$ \\ 
z3 & -6.37 & -9.17 & 0.54 & 1.26 & 0.97& $0.27\pm0.11$   &  $-$ & $-$ & $176\pm74$ \\ 
z4 & -5.81 & 6.79 & 0.42 & 1.24 & 1.09& $0.30\pm0.11$   &  $-$ & $-$ & $322\pm119$ \\ 
z5 & -4.55 & -3.01 & 0.55 & 1.26 & 0.95& $0.25\pm0.11$   &  $-$ & $-$ & $158\pm70$ \\ 
z6 & $-0.21\pm0.06$ & $-0.09\pm0.07$ & 0.54 & 1.26 & 1.81& $1.71\pm0.26$   & $0.59\pm0.08$ & $50\pm5$ & $245\pm37$ \\ 
z7 & 2.73 & -7.77 & 0.43 & 1.24 & 1.04& $0.27\pm0.11$   &  $-$ & $-$ & $256\pm103$ \\ 
z8 & 5.39 & 10.85 & 0.34 & 1.22 & 0.82& $0.12\pm0.07$   &  $-$ & $-$ & $422\pm237$ \\ 
z9 & 7.91 & 3.15 & 0.56 & 1.27 & 0.86& $0.19\pm0.10$   &  $-$ & $-$ & $539\pm273$ 
\enddata
\tablecomments{
For the main cluster component ($z_6$), the halo centroid was allowed to vary, and an eNFW model was fitted. The concentration parameter was also fitted here, whereas for all other less massive halos the concentration was set by the mass-concentration relation given by \cite{Duffy08}.
\tablenotetext{a}{The centroid position of each halo is given relative to the cluster center (see Table~\ref{tab:cluster}) in units of 
arcminutes.}
\tablenotetext{b}{The median photometric redshift for each halo estimated from the green sample.}
\tablenotetext{c}{The photometric source redshift for each halo.  In the multi-halo shear fitting process, we assumed a flat prior over the range $z_s\pm 0.1$.}
\tablenotetext{d}{Marginalized bi-weight center and scale locations are reported}
\tablenotetext{e}{The projected mass ellipticity of our eNFW model is defined as $e=1-b/a<1$ with $a$ and $b$ the major and minor axes of the isodensity contours.}
\tablenotetext{f}{The position angle of the eNFW halo major axis is given in units of degrees, measured north of west.}
}
\end{deluxetable}

%


\end{document}